\documentclass{article}

\usepackage{cite}
\usepackage{amsmath,amssymb,amsfonts}
\usepackage{graphicx}
\usepackage{textcomp}
\usepackage{xcolor}
\usepackage{color}

\def\BibTeX{{\rm B\kern-.05em{\sc i\kern-.025em b}\kern-.08em
    T\kern-.1667em\lower.7ex\hbox{E}\kern-.125emX}}

\usepackage[utf8]{inputenc}
\usepackage{comment}

\usepackage[noend]{algpseudocode}
\usepackage{algorithm}

\newtheorem{definition}{\textbf{Definition}}[section]

\usepackage{geometry}
\usepackage{pgf,tikz,pgfplots}
\pgfplotsset{compat=1.15}
\usepackage{mathrsfs}
\usetikzlibrary{arrows}

\usepackage{textcomp}

\usepackage{longtable}
\usepackage{array} 
\DeclareMathOperator*{\argmin}{arg\!\min}
\DeclareMathOperator*{\average}{avg}

\usepackage{authblk}
\usepackage{hyperref}

\title{A Geometric Approach to Passive Localisation}
\author[1]{\href{mailto:Theofilos.Triommatis@liverpool.ac.uk}{Theofilos Triommatis} }
\author[1]{\href{mailto:potapov@liverpool.ac.uk}{Igor Potapov}}
\author[2]{\href{mailto:Gareth.Rees@mbda-systems.com}{Gareth Rees}}
\author[1]{\href{mailto:jfralph@liverpool.ac.uk}{Jason F. Ralph}}

\affil[1]{
School of Electrical Engineering and Electronics and Computer Science
\textit{University of Liverpool}
Liverpool, L69-3BX, UK }
\affil[2]{\textit{MBDA} Bristol, BS34 7QS, UK}

\providecommand{\keywords}[1]
{
  \small	
  \textbf{\textit{Keywords---}} #1
}

\begin{document}

%\textregistered \textcopyright 
%\sffamily\textregistered\textcopyright

\date{}

\maketitle

\begin{abstract}
In this paper, we present a geometric framework for the passive localisation of static emitters. The objective is to localise the position of the emitters in a given area by centralised coordination of mobile passive sensors. This framework uses only the geometry of the problem to minimise the maximal bounds of the emitters' locations without using a belief or probability distribution. This geometric approach provides effective boundaries on the emitters' position. It can also be useful in evaluating different decision-making strategies for coordinating mobile passive sensors and complementing statistical methods during the initialisation process. The effectiveness of the geometric approach is shown by designing and evaluating a greedy decision-making strategy, where a sensor selects its future position by minimising the maximum uncertainty on its next measurement using one of the global objective functions. Finally, we analyse and discuss the emergent behaviour and robustness of the proposed algorithms.
\end{abstract}


\section{Introduction}

 The problem of localising emitters inside a given area using passive sensors is of great interest across a range of applications of autonomous vehicles in air~\cite{dogancay2012uav,7040040_UAV1}, space~\cite{DBLP:journals/automatica/HuSC21_space}, and underwater~\cite{8756257_uwater1,s21134457_uwater2}.
 %In this paper, we consider the case where emitters are static
 In this paper, we consider the problem with static emitters and mobile sensors. We introduce a novel geometric approach to study the exploration process and develop effective decision-making mechanisms.
 %that lead to an optimal trajectory for a given objective.
 %\textcolor{orange}{Theo: this is our aim for the framework to find strategies that are optimal even if the greedy ones may be not one of them.} 
 Due to uncertainties in received signals as well as an infinite number of 
 possible trajectories in continuous space or an exponential one in a discrete space, 
 the brute-force solution
 is computationally infeasible.
% It is computationally infeasible to examine all possible trajectories. 
To overcome this challenge, we consider geometric objectives that can guarantee the solution's quality by limiting in the worst case scenarios and by replacing the continuous dynamics with a sequence of  discrete steps in
a limited number of directions for mobile  sensors.
%
%In an attempt to approximate an optimal one, we divide time into time steps and we consider situations where sensors are allowed to move in a limited number of directions. This makes the problem tractable and means that the trajectories computed are sequences of discrete steps.

Passive sensors only receive signals and provide an angle (with an error) which indicates the emitter's direction.
Since the sensor's measurements are not precise the direction is given 
 by a cone centred around its line of sight and this region is assumed to contain the emitters. 
%In this settings, a sensor's measurements correspond to a cone centred around its line of sight and this region is assumed to contain the emitters' positions. 
By considering a 2D version of the problem  the cone is reduced to a triangle.
%We also restrict the emitters' locations to a 2D horizontal plane, meaning that the cone reduces to a triangle. 
The whole localization process is iterative with a sensor detecting the presence of an emitter within an area, and calculating the intersection of the measurement with the polygon that bounds currently the emitter's location. The decision-making happens when a sensor picks a direction to move to the next position to get the next measurement.
%, which depends on the observed polygons up to this point. 
%We use a greedy algorithm that minimises the maximum bounded area in the next step by evaluating one of a set of global objective functions.
 
The main contribution of this work is to develop an approach that is based purely on the geometry of the search task; an approach that can complement more sophisticated methods based on Bayesian inference by bounding regions being considered for further processing. In conventional approaches, the uncertainties in the emitter locations are treated as a probabilistic problem. Often, minimising uncertainties involves Markov decision processes to create decision-making mechanisms with various metrics~\cite{sarkka_2013,ZhaoWWCS19,10.1117/12.817315/metrics}. One example is the determinant of the covariance matrix, which can be used to define the area of an ellipse centred on the emitter's estimated position~\cite{DBLP:journals/taes/YangKB12_cov_mat1,DBLP:conf/nma/MihaylovaLBGS02_cov_mat2}. 
Our geometric approach localises the emitters by  polygons instead of ellipses. We aim to minimise the worst-case polygonal area (or areas) over the routes allowed to the sensor and a
set of  global objective functions.
%In this approach presented here, instead of ellipses, we have polygons and we study objective functions that minimise the worst-case polygonal area (or areas) over the routes allowed to the sensor. 
%This framework evaluates such decision-making mechanisms by introducing a set of intuitive global objective functions.

In particular, this geometrical approach can be useful at the time when emitters have been detected for the first time. At such a moment it is hard to make accurate predictions on the emitters' positions because there is not enough gathered
 data to form a reliable probability distribution or belief.
%During these periods of time, the data gathered are not enough to make accurate predictions on the emitters' position based on a prior probability distribution or belief. 
In other words, the bounds on the estimates can be very large, which increases the probability of error. 
%In particular, this framework can help in periods where emitter locations are being initialised because 
%This framework
Our approach provides reliable polygonal bounds on the location of an emitter by only considering the geometry of the problem without using a prior distribution -- thereby reducing the reliance on initial assumptions and initial data. 
The geometric approach is in line with other related attempts providing upper and lower bounds for estimates without using a probabilistic approach e.g. the algebra of intervals~\cite{DBLP:journals/jucs/Markov98/Alg_interv}. It is also somewhat related to probabilistic approaches using random finite sets~\cite{DBLP:journals/taes/GaoBCF21/RFS}. However, our approach provides purely geometric bounds and does not involve expensive calculations.

We begin in section~\ref{sec:Framework} by introducing the framework for the passive localisation of static emitters. In section~\ref{subsec:Framework_Setting_Formulation}, we formally present the problem of localising emitters with a passive sensor and particular aspects of the problem that affect the decision-making of the sensors. In sections~\ref{subseq:One_S_One_E} to \ref{subseq:m_S_k_E} we define mathematical tools to address four cases (one-to-one, one-to-many, many-to-one and many-to-many) that depend on the number of emitters and the number of the sensors. 
%Using the mathematical tools defined in section~\ref{sec:Framework} we present, 
In section~\ref{sec:Experiments} we present the outcome of experiments and  study the stability of the greedy algorithms according to the noise,  their emergent behaviour, and the accuracy of the algorithms when number of step directions is varied. Finally, in section~\ref{sec:Discuss}, we summarise the results and draw our conclusions from the experiments and the framework. 

\section{Framework of Exploration}
\label{sec:Framework}

\noindent
In this section, we design a novel framework to examine how passive sensors explore an area by bounding the locations of emitters that the area contains. This framework is used to analyse how short-term decisions affect the path of the exploration, and to evaluate different policies that the sensors could follow to localise the emitters. Most studies in this area, because of its probabilistic nature, often focus on methods to determine sensor paths based around directed random walks, either Brownian in nature or L\'{e}vy flights~\cite{Levi_Random_Search_Pattern}.  In our approach, statistical factors to make predictions for the emitters' positions are not used. 
We show that geometric constraints on their own could provide effective decision-making procedures for the localisation problem.

%This means that only the geometry of the problem is taken into consideration in the decision-making process.

\subsection{Problem Formulation}
\label{subsec:Framework_Setting_Formulation}
Let us define a model with $k$ static emitters and $m$ mobile passive sensors. Sensors are imperfect and can only provide a range of angles from which a signal may originate and, because they have limited range, a measurement is represented by a cone that contains the position of the source (or sources) of the transmissions (the emitters). 
This compares with the dominant approach in the literature where each measurement is represented as a random variable~\cite{dogancay2012uav,7040040_UAV1,DBLP:journals/automatica/HuSC21_space,8756257_uwater1,s21134457_uwater2,Gustafsson489131 }. By assuming the measurement to fall within a cone we only need to consider the case where the measurements provide bounds
%By considering the measurement to fall within a cone, we assume a pessimistic case where the measurements provide bounds 
on a location without specifying an associated probability distribution. However, although this problem can be addressed in 3D, for simplicity, we restrict the emitters locations to a 2D plane. This means that the sensor measurements form a triangle that contains the source of transmission rather than a cone. So a measurement will provide a triangular area, rather than a single angle value (with an associated error). In this paper, and without loss of generality, we assume that a sensor has sufficient flexibility and sensitivity that it has a field of regard of $360^o$ and a range that goes beyond the limits of the investigated area. We also use a centralised approach to decision making -- that is, the information that the sensors gather are instantly shared, so every sensor stores the global picture -- but other alternative approaches are also discussed.

The starting position of the $i$-th sensor can be a point in the plane $x_0^{(i)} \in \mathbb{R}^2$. As the sensor moves, the challenge is to select the next position to take the next measurement of the emitters. In a continuous setting, every point inside the circle centred on the sensor's current position, with a radius that depends on the maximum speed of the sensor, should be taken into consideration. We approximate infinite number of positions on the circle by a finite set of  equally distributed discrete points to define neighboring positions as one of the parameters. For example, a special case of a
hexagonal tessellation for navigation (i.e. considering 6 points) has been recently used in~\cite{honeycomb}.

%But, examining every point inside the circle is computationally infeasible. Inspired by~\cite{honeycomb} where the authors use hexagons to navigate, we consider discrete points on a circle, with a predefined radius to define neighboring positions.

\begin{definition}[Sensors' Discrete Directions]
Given a radius $r$, and an integer $k>1$ then the circle centered on $(0,0)$ with radius $r$ is divided to $k$ points $d_1,\ldots,d_k$, where $d_1,\ldots,d_k$ are the vertices of a regular $k$-gon. The set of sensors' discrete directions $\mathcal{D}$ is comprised from the vectors $d_1,\ldots,d_k$, which yield the next position when added to the current one.
\end{definition}

Note that the trajectory of the $i$-th sensor depends on its current position  $x^{(i)}$ and  available $k$ directions. Thus, the optimal continuous trajectory could be approximated by increasing the number of possible directions.

%Finally, choosing the next move from discrete points on a circle leads to an approximate solution rather than an optimal trajectory.
 
\begin{definition}[Trajectory]
The trajectory or path is a sequence of consecutive positions of the sensor, that is, $\{x_1,\ldots,x_k \} \equiv x_{1:k}$, where $x_i \in \mathbb{R}^2$
\end{definition}

 With the first measurements, the sensors form triangles around the emitters, but as the exploration continues, they become polygons because each measurement intersects with the `observed area'. The observed area is the area that the sensors can assume to contain the emitters. The term `observed area' is used because there can be cases where two emitters are in one polygon and falsely perceived as one emitter. 
 To give an example of such a situation: the sensor and the two emitters are in co-linear positions, and the emitters are in the same direction, so the sensor will perceive the transmissions from the two emitters as one, forming one triangle that contains both of them. 
 In other words, the sensors assume that there are at least as many emitters as the polygons observed. We evaluate the observed area by measuring the area of their polygons. We note that the problem of associating measurements to the proper emitter has been the subject of extensive study~\cite{DBLP:journals/pami/CoxH96/assoc1,DBLP:journals/inffus/GruyerDMB16/assoc_DS,DBLP:journals/access/ZhaoWWMG21/assoc3,DBLP:journals/access/LeeJK21b/assoc4,DBLP:journals/eswa/RakaiSSZY22/aasoc5} but for the moment, we study the case where there are no mis-associations.
 
 Time-related issues in the problem can be handled by assuming that a universal clock exists.  
 %Even if it does not, one of the sensors can keep track of time, to provide a stable time reference.  
 Using this universal clock, time can be divided into time slots which we will call time steps.
 It is assumed that in each time step, the emitters send a transmission.
 With this framework, we will analyse the trajectories $x_{1:T}$ that are generated, and evaluate the effect of different policies on the sensors' decisions.
 To summarise, in each time step the following events happen:
 \begin{enumerate}
     \item The emitters send a transmission;
     \item The sensors receive those transmissions;
     \item The sensors compute the intersection of the measurements with the observed area;
     \item The sensors evaluate 
     %determine where to move next 
     neighboring positions 
      using an objective function that depends on the sensors' positions and  observed area; 
     \item The sensors move to the most optimal next position.
 \end{enumerate}
 
At the moment we consider centralised coordination procedure,  where all sensors move synchronously in every time step to achieve a better localisation across all sensors.
%
%of the sensors move during each time step to achieve a better localisation across all sensors.
%The problem is a centralised scenario is where all of the sensors move during each time step to achieve a better localisation across all sensors. 
To be more precise, the problem lies with the selection of an objective function that makes short-term decisions that lead to an efficient localisation overall.   
Table~\ref{tab:Table_aspects_frame} summarises all the assumptions made.

 \begin{table}[htbp]
\begin{centering}
    \begin{tabular}{|l|l|}
    %\begin{tabular}{|p{2.6cm}| p{2.6cm}| p{2.6cm}| p{2.6cm}|p{2cm}|}
    %\begin{table}{||L|L||L|L||}
    \hline
    \hline
    \label{table_Approaches_Framework}
    \textbf{Problem's Aspect} & Approach 
    %& \textbf{Parameters} & Approach  
    \\
    \hline
    \hline
    \textcolor{blue}{\textbf{Sensor's possible directions:}} & Given Points on a Circle 
    %& Relocatable & Moving Emitters &
    %Neighborhood also referrence
    \\
    \hline
    \textcolor{blue}{\textbf{Number of Sensors:}} &  Single / multiple 
    %Euclidean distance error & & & 
    \\
    \hline
    \textcolor{blue}{\textbf{Sensors' Starting Position:}} & Different starting positions
    %& \small{Same Starting Position} & \small{Different starting positions} & &   
    %& Unknown & & 
    \\
    \hline
    \textcolor{blue}{\textbf{Sensors' Uniformity:}} & Same type of Sensors
    %\small{Different types of sensors} & & 
    %& Local Point to Point & Cluster Broadcast & &
    \\
    \hline
    %\textcolor{Blue}{\textbf{Field of View:}} & \small{Few Degrees (Usually $1^o$ - $5^o$)} & & & 
    %\\[2ex]
    %\hline
    \textcolor{blue}{\textbf{Field of Regard:}} & $360^o$ and Inf Range 
    %& $360^o$ and 200 km Range & &
    %Centralised Shared Memory & Centralised Hierarchy of Memory & Distributed Hierarchy of Memory &
    \\
    \hline
    %\textcolor{Blue}{\textbf{Stochastic Model:}} & Bayesian Approach & Fuzzy Logic & Dempster – Shafer theory & Probabilistic State Space
    %\\[2ex]
    %\hline
    %\textcolor{Blue}{\textbf{Filters:}} & Kalman Filter & Extended Kalman Filter & Particle Filter & 
    %\\[2ex]
    %\hline
    %\textcolor{Blue}{\textbf{Filtering Techniques:}} & Passive - Ranging & Pseudo - Ranging & & 
    %\\[2ex]
    %\hline
    \textcolor{blue}{\textbf{Number of emitters:}} & 
    Unknown to the sensors
    %Known a priori 
    %Centralised - Cooperative & \small{Distributed - Cooperative Locally by clusters} &  & 
    %& Unknown & &
    \\
    \hline
    \textcolor{blue}{\textbf{Emitters' Motion State:}} & Static 
    \\
    \hline
    \textcolor{blue}{\textbf{Emitters' Location Error:}} & Area of Polygon
    \\
    \hline
    \textcolor{blue}{\textbf{Frequency of transmissions:}} & Fixed (1 per time-step)
    \\
     & \textit{(A Simplification for this work)}
    \\
    \hline
    \textcolor{blue}{\textbf{\small{Communications}}} & 
    Global 
    \\
    \hline
    \textcolor{blue}{\textbf{Information Storage:}} & Centralised Shared Memory%Distributed Hierarchy of Memory
    \\
    \hline
    \textcolor{blue}{\textbf{Decision Making:}} & 
    Distributed - Coordinated 
    \\
    \hline
   \hline
    %\end{longtable}
    \end{tabular}
    \caption{The aspects of the framework}
    \label{tab:Table_aspects_frame}
\end{centering}
\end{table}

%Moreover, in each time step a sensor observes an area of a polygon $A(x_{1:t})$ that contains the emitter, where $x_t\in \mathbb{R}^2$ is the position of the sensor at time-step $t\in \mathbb{N}$ and $x_{1:t}$ is the path $x_1,\ldots,x_t$.

We will analyse four distinct cases in terms of (number of sensors, numbers of emitters). The cases are the following: $(1,1),(1,k),(m,1),(m,k)$. Next we describe each of the scenarios and introduce overall objectives.

\subsection{One Sensor - One Emitter}
\label{subseq:One_S_One_E}

\begin{definition}[Observed Area of the Individual]
\label{Def:Observed_Area_11}
The observed area of the individual is defined inductively, as follows, with the first measurement $P_1 = M_1$. 
Let $P_{t-1}$ be the polygon area that the sensor had observed to contain the emitter in the past, and $M_t$ be the current measurement of a transmission. The new observed area of the individual sensor is $P_t$, where
\begin{align}
    P_t = P_{t-1}\cap M_{t} = \bigcap_{i=1}^t M_{i}
\end{align}
It is apparent that the observed area depends on the trajectory that has been followed up to the current time step, so equivalently, the notation $P(x_{1:t})$ instead of $P_t$ will be used.
\end{definition}
\vspace{0.1cm}

In this scenario, we have a set of available directions $\mathcal{D}$ and a family of objective functions $\mathcal{F}$ these function present the criteria of a short-term decision. The functions taken into consideration are the following:
\begin{itemize}
    \item Given the current position $x$, and the observed area $P$, minimise the maximum area $A$ 
    %https://www.overleaf.com/project/6139c9b657f722dbb6519647a 
    of the intersection between the observed area and the triangle of the sector's field of view 
    \begin{align}
    \label{eq:Obj_area}
        f_1(P,x) = \min_{d\in \mathcal{D}}{
        \left(
        \max_{\theta \in [0,2\pi]}{A(\theta,x+d)}
        \right)
    }
    \end{align}
    In this case, the objective is to minimise the uncertainty of the area that localises the emitter, in case the sensor moves towards the direction $d^*\in \mathcal{D}$ such that
    \begin{align}
        d^*=\argmin_{d\in \mathcal{D}}{
        \left(
        \max_{\theta \in [0,2\pi]}{A(\theta,x+d)}
        \right)
        }
    \end{align}
    \item Given the current position $x$, and the observed area $P$, minimise the diameter of the maximum area of the intersection $A$ between the sensor's field of view with direction $\theta$ and $P$
    \end{itemize}
    \begin{align}
    \label{eq:Obj_diam}
        f_2(P,x) = \min_{d\in \mathcal{D}}{
        \left( diameter \left(
        \max_{\theta \in [0,2\pi]}{A(\theta,x+d)}
        \right)
        \right)
    }
    \end{align}
    \begin{itemize}
    \item[]
    This case could help to distinguish between stretched polygons with small area and more evenly centred polygons.
\end{itemize}

\noindent
We must note that it is crucial to be able to compute the maximum intersection between a polygon and a measurement that has direction $\theta$. Moreover, notice that $\mathcal{D}$ is discrete, but the measurement's direction is the continuous interval $[0,2\pi]$.

\begin{figure}[ht]
    \centering
    \begin{tabular}{c c}
        \scalebox{0.104}{
        \includegraphics{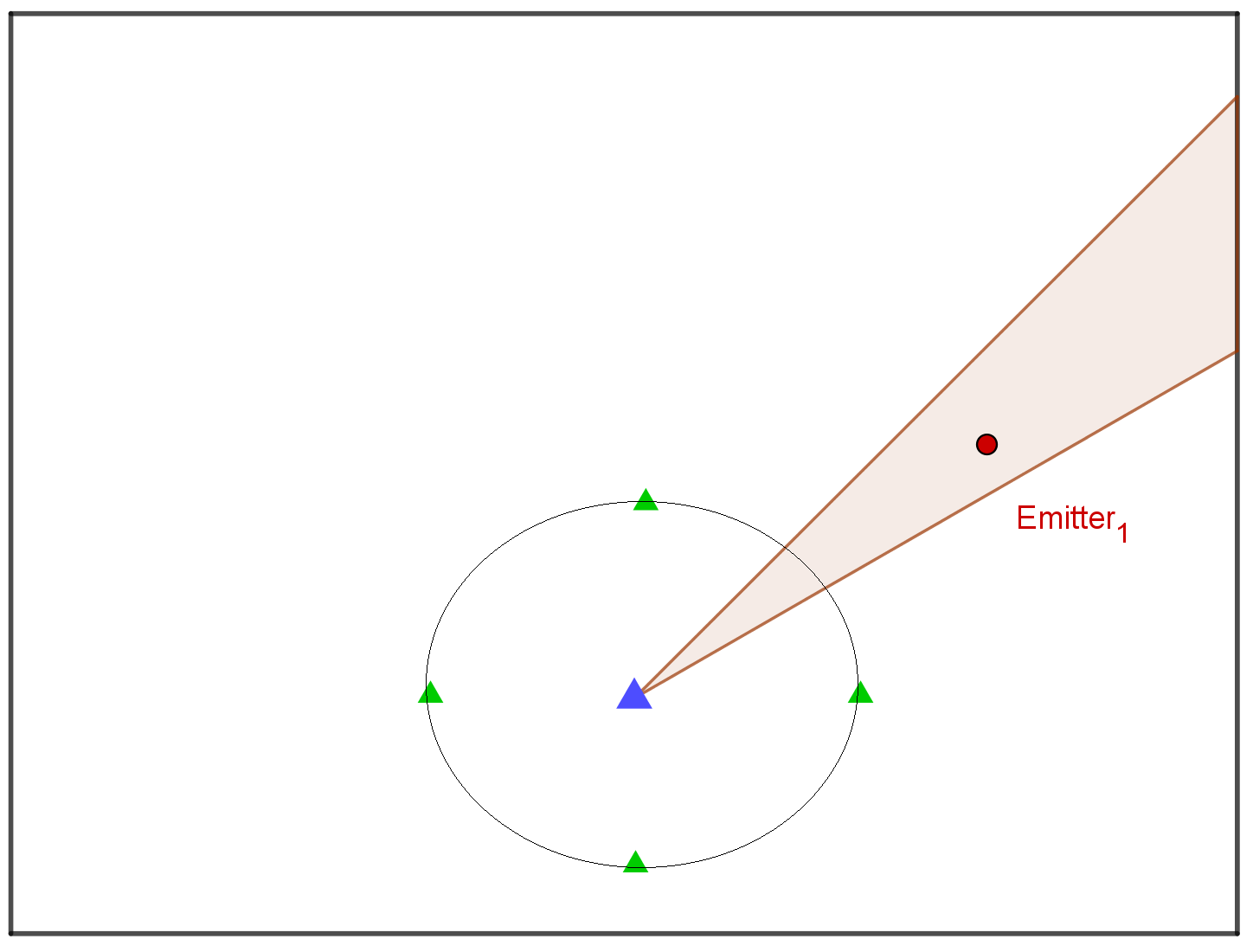}}
         &
        \scalebox{0.104}{
        \includegraphics{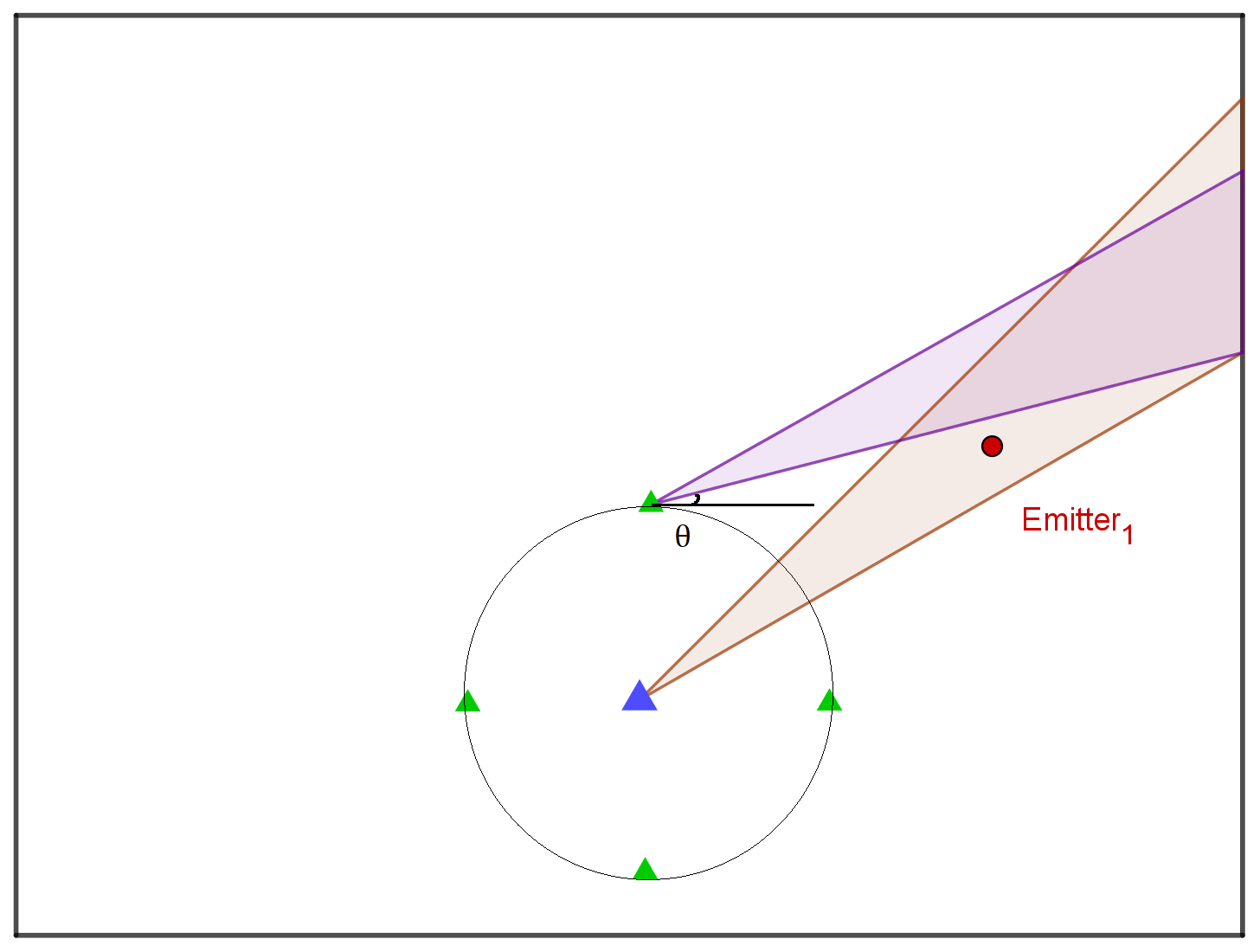}}
    \end{tabular}
    
    \caption{In this example the sensor is allowed to travel only North, East, South or West for this exploration. In this case the set of available directions is $\mathcal{D} = \left\{(1,0),(0,1),(-1,0),(0,-1) \right\}$. For every $d_i \in \mathcal{D}$ the sensor computes the maximum intersection that depends on $\theta \in [0,2\pi]$ according to equations~(\ref{eq:Obj_area}), (\ref{eq:Obj_diam}). }
    \label{fig:move_directionl}
\end{figure}

Apart from $\mathcal{D}$ and $\mathcal{F}$, a set of conditions for termination of the exploration $\mathcal{T}$, and a set of constraints $\mathcal{C}$ are introduced. The set of constraints should express constraints of the sensor's movement. The set conditions for the exploration's termination is self-explanatory, and in our scenario, these could be one of the following two:
\begin{enumerate}
    \item The exploration stops after a given amount of time steps by providing an upper bound $t_{max}$. So the condition $C$ is to check if the current time step is greater than the given $t_{max}$, that is
    \begin{align}
        t_{current}> t_{max}
    \end{align}
    \item The exploration can also stop when the emitter is localised, up to a given tolerance. Specifically, at time step $k$, when the area $Area(P_k)$ of the observed area is small enough, the exploration stops, that is
    \begin{align}
        Area(P_k) \leq tol &&  \mbox{where }tol \mbox{ is the tolerance }
    \end{align}
\end{enumerate}

\begin{algorithm}[H]
\caption{\textit{Greedy Algorithm Framework of Local Exploration}}
\label{AlgCaseFrame}
\begin{algorithmic}[1]
  \Require { Initial position $x_0\in\mathbb{R}^2$, a constraint $C \in \mathcal{C}$, the set of sensors' possible directions $\mathcal{D}$,
  a termination condition $fin\in\mathcal{T}$, an objective function $f\in\mathcal{F}$}
  \Ensure The trajectory $t=x_{0:k}$ of the sensor.
  \State $k=1$
  \While{ $fin$ is \textbf{false}}
  \State \textbf{Get} the measurement of the transmission
  \State \textbf{Compute} the intersection of the measurement and the observed area
  \For{ every $d_i \in \mathcal{D}$}
  \State $p_i = f(x_k+d_i) $  //Evaluate the Objective function
  \EndFor
  \State $d^* = \argmin(p)$ // Find the direction that minimises the objective function
  \State $x_k = x_{k-1}+d^*$ // Move to the "best" position 
  \State $k = k+1$
  \EndWhile
  \State \Return $x_{0:k}$
\end{algorithmic}
\end{algorithm}

\noindent
Finally, to define the optimal solution,  we introduce the set of global objectives $\mathcal{GO}$. This set provides the goal of the exploration, an example is the following objective:
\begin{enumerate}
    \item When the stopping condition is “stop when reach the maximum time steps bound $t_{max}$ ”, then an objective is to minimise the area observed when $t_{max}$ is reached:
    \begin{align}
    \label{Eq:Global_11_Min_area}
        \min_{x_{1:t_{max}}}Area(P\left(x_{x_1:t_{max}}) \right)
    \end{align}
    \item  When the stopping condition is “stop when the observed area is smaller than a tolerance value $tol$”, then an objective is the following: Let $Tr$ be the set of the trajectories which in the end achieve the observed area to be less than $tol$, that is
    \end{enumerate}
    \begin{align}
        &Tr=\\
        &\left\{ x:\mathbb{N}\longrightarrow \mathbb{R}^2 \mid  \exists\ T \mbox{ such that } Area(P(x_{1:T}))<tol \right\}
        \nonumber
    \end{align}
    \begin{enumerate}
    \item[]
    Find the $x^*_{1:t_0}$ that the minimum length, that is
    \begin{align}
    \label{Eq:Global_11_Min_Steps}
        t_0 = \min_{x_{1:t}\in Tr} \left\{t\in \mathbb{N}: Area(P(x_{1:t}))<tol \right\}
    \end{align}
\end{enumerate}

\subsection{One Sensor - \textit{k} Emitters}

\begin{figure}[ht]
    \centering
    \scalebox{0.13}{
    \includegraphics{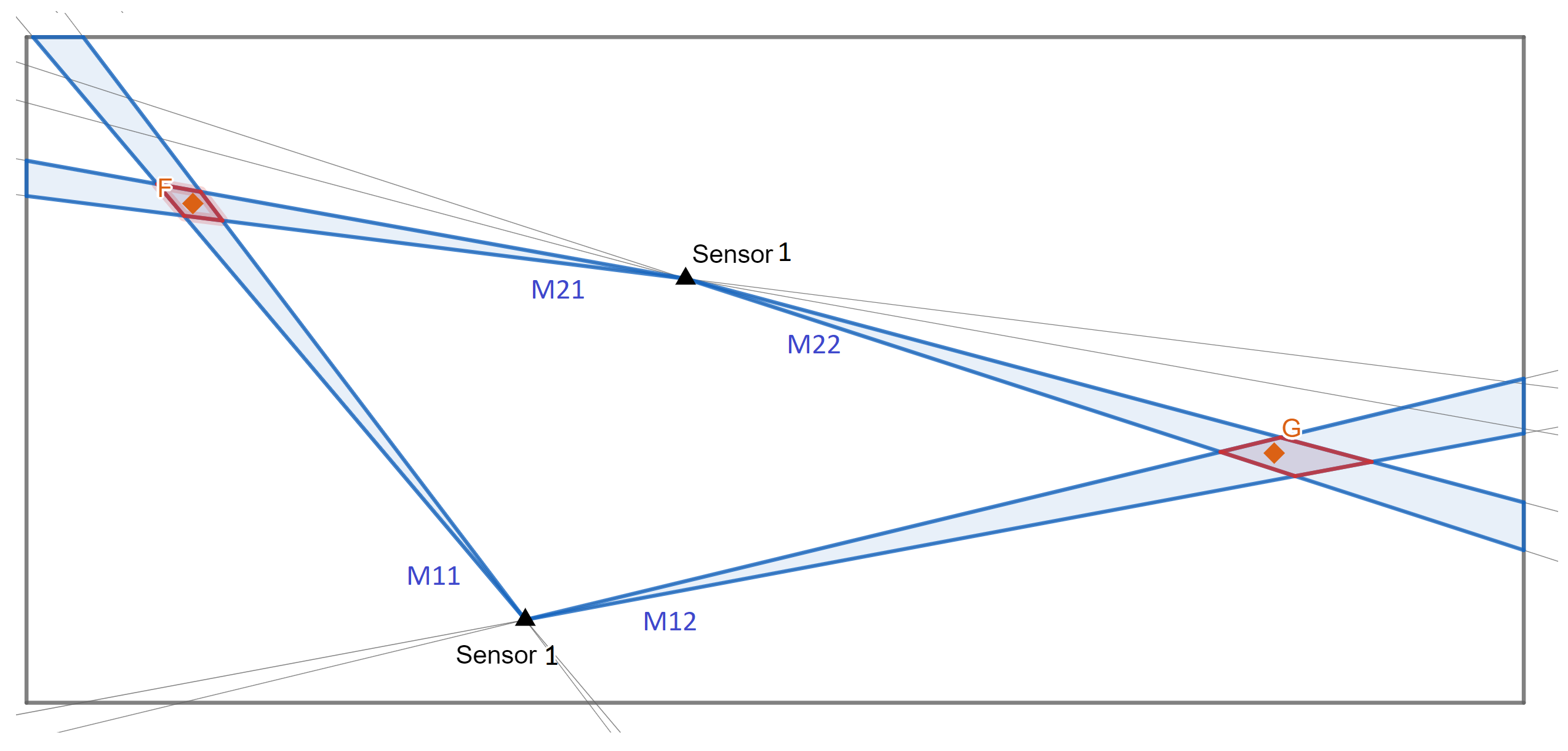}}
    \caption{These are two time steps as sensor 1 explores by moving from south to the north. Note that the observed area in the first time step is $P_1=M_{11}\cup M_{12}$, and in the second time step $P_2= (M_{11}\cap M_{21}) \cup (M_{12}\cap M_{22}) \equiv  F \cup G$, with $P_{21}=F$ and $P_{22}=G$.}
    \label{fig:1S2E}
\end{figure}

\begin{definition}[Observed Area of the Individual]
\label{def:Obs_Area_k1}
Let $\boldsymbol{M}_{i}= (M_{i1},\ldots,M_{i\ell(i)})$, where $\ell(i)\in \{1,\ldots,k\}$ is the number of observed emitters in the $i$-th measurement that the sensor gets. The observed area of the individual sensor is defined inductively as follows:
\begin{align}
    P_1 = \bigcup_{j=1}^{\ell(1)} M_{1j} 
\end{align}
Let $P_{t-1}$ be the area that the sensor had observed to contain emitters in the past, and $\boldsymbol{M}_{t}$ be the current measurements of a transmission. The new observed area of the individual sensor is $P_t$, where
\begin{align}
    P_t = \bigcup_{j=1}^{\ell(t)} \left( P_{t-1}\bigcap M_{tj} \right)
\end{align}
In this case the observed area can be interpreted as a union of polygons so with the notation $P_{ti}$ we shall refer to the observed area at time step $t$ and to the $i$-th polygon.
\end{definition}

In this scenario we still have the family of objective functions $\mathcal{F}$ and the sets $\mathcal{C}$,$ \mathcal{D}$, $\mathcal{T}$ as in section~\ref{subseq:One_S_One_E}. Note that in family $\mathcal{F}$ there will be more complex objective functions, for example:
Given the current position $x_t$, and the observed area $P_t$, minimise the total area of the maximum intersections $A_i$, $i \in \{1,\ldots,\ell(t)\}$, of the observed area and the triangle of the sector's field of view 
    \begin{align}
    \label{eq:Obj_area_1S_kE}
        f_1(P,x) = \min_{d\in \mathcal{D}}{
        \left(
        \max_{\theta_i \in [0,2\pi]}{\sum_{i=1}^{\ell(t)}A_i(\theta_i,x+d)}
        \right)
    }
    \end{align}
    Function $A_i(\theta,x)$ represents the area of intersection between the polygon $P_{ti}$ with the sensor's field of view which is centered at position $x$ and has direction $\theta$.
Another one is the following: Given the current position $x_t$, and the observed area $P_t$, minimise the largest observed polygon $P_{tj}$, $i\in \{1,\ldots,\ell(t) \}$
\begin{align}
    \label{eq:Obj_minmax_1S_kE}
    \hspace{-0.2cm}
        f_2(P,x) = \min_{d\in \mathcal{D}}{
        \left(
        \max_{i \in \{1,\ldots,\ell(t) \}}
        \left(
        \max_{\theta_i \in [0,2\pi]}{A_i(\theta_i,x+d)}
        \right)
        \right)
        }
\end{align}

To solve the problem for multiple emitters, we need to introduce a set of global strategy $\mathcal{GS}$ to achieve global objectives in $\mathcal{GO}$.
In this scenario, there are many emitters, and this can be reflected at $\mathcal{GO}$ as well.
\begin{enumerate}
\item 
  Minimise the area observed when $t_{max}$ is reached
    \begin{enumerate}
        \item Minimise the total area
        \begin{align}
        \label{eq:Go_max_step}
            \min_{x_{1:t_{max}}}Area\left(P(x_{x_1:t_{max}}) \right)
        \end{align}
        \item Minimise the area of the maximum polygon observed
        \begin{align}
        \label{eq:Go_tol_areap}
            \min_{x_{1:t_{max}}}Area\left(\max_{i}P_i\left(x_{x_1:t_{max}}\right) \right)
        \end{align}
    \end{enumerate}
    \end{enumerate}
    %\item 
For the second stopping condition taken into consideration, the same can be applied when the set of trajectories $Tr$ is defined. For example, an objective is to get the average area of the polygons to be less than a given value $tol$.
Let $Tr$ be the set of the trajectories which achieve in the end the average observed area to be less than $tol$ that is
\begin{align}\hspace{-0.2cm}
    Tr=
    \left\{ x:\mathbb{N}\rightarrow \mathbb{R}^2 \mid  \exists\ T \mbox{ s.t.}\average_i
    \left(
    Area(P_i(x_{1:T}))<tol 
    \right)
    \right\} 
    %\nonumber
\end{align}

\begin{enumerate}
    \setcounter{enumi}{1}
    \item Find the $x^*_{1:t_0}$ that the minimum length, that is
    \begin{align} \hspace{-0.3cm}
        x^*_{1:t_0} = \argmin_{x_{1:t}\in Tr} \left\{t\in \mathbb{N}: \average_{i}
        \left(
        Area(P_i(x_{1:t}))<tol
        \right)
        \right\}
    \end{align}
\end{enumerate}
Note that Algorithm~\ref{AlgCaseFrame} will work  perfectly in this scenario, the only thing that distinguishes between the two scenarios is just the input of the algorithm.

A global strategy taken into consideration is the following:
Once the sensor gets the first measurements $\boldsymbol{M}_{1}$, use Algorithm~\ref{AlgCaseFrame} with the objective function~(\ref{eq:Obj_area}), which means the sensors makes decisions only for $M_{1i}$, $i \in \{1,\ldots,\ell(1)\}$ for $T_1>0$ time steps, while still perceives all the transmissions, and then follow another measurement $M_{T_{1}i}$, $i \in \{1,\ldots,\ell(T_1)\}$. In this policy the challenge is to find a way choosing combinations of steps and areas to move towards to to achieve an efficient localisation of the emitters.

\subsection{\textit{m} Sensors - One Emitter}
\begin{definition}[Global Observed Area]
\label{def:Obs_Area_m1}
The global observed Area $P$, is defined inductively using the observed area of the individual sensor $i$, notated with $P^{(i)}$, as follows:
\begin{align} 
     P_1 = \bigcap_{i=1}^{m}P_1^{(i)} = \bigcap_{j=1}^{m}M_{1}^{(j)}
\end{align}
Let $P_{t-1}$ be the global observed area at time step $t-1$, and $M^{(j)}_t$ be the current measurement of the $j$-th sensor. The global observed area at $t$, is $P_t$, where
\begin{align} \hspace{-0.3cm}
    P_t = \bigcap_{i=1}^{m}P_t^{(i)} = \bigcap_{j=1}^{m} \left( P_{t-1}\cap   M^{(j)}_{t} \right) = \bigcap_{i=1}^t \left(  \bigcap_{j=1}^{m}M_{i}^{(j)} \right) 
\end{align}
\end{definition}

In this scenario we still have the set of global objectives $\mathcal{GO}$ and the sets $\mathcal{C}$,$ \mathcal{D}$, $\mathcal{T}$ as in section~\ref{subseq:One_S_One_E}.
A short-term objective in $\mathcal{F}$ can be the minimisation of the intersection between sensors' minmax(area) as defined in equation~(\ref{eq:Obj_area}). Moreover, it should be clarified that with m Sensors the following actions happen in every time step:
 \begin{enumerate}
     \item The emitters send a transmission
     \item The sensors receive those transmissions
     \item The sensors compute the intersection of the measurements with the observed area
     \item The sensors determine where to move next by: 
     \begin{enumerate}
         \item Evaluating their objective function 
         \item Sharing the information gathered
         \item Making a decision where to move next
     \end{enumerate}
     \item The sensors move to the next position
 \end{enumerate}
A major difference between this case and the previous ones is that the global picture is constructed from the sensors' partial information. There are three different categories of information sharing~\cite{liggins1997distributed}. In the first one, every sensor has all the information. In the second one, the information is forwarded to a subset of sensors, and in the final one, there is no information sharing at all. As mentioned, a centralised approach is considered here, which means that every sensor will have the global picture. A related problem is `gossiping' of the sensors, which has been studied in similar cases~\cite{Gossip_Radio_CN_20, Odd_Gossip}. 
Finally, this case can be thought the same as in section~\ref{subseq:One_S_One_E} with the additional fact that with many sensors, there is a parallel algorithm to achieve the objective $O \in \mathcal{GO}$. 

\subsection{\textit{m} Sensors - \textit{k} Emitters}
\label{subseq:m_S_k_E}
We will use and generalise all these definitions to study the most general case of $m$ Sensors and $k$ Emitters and get results. In this scenario, we will define the observed area by combining the definitions~\ref{def:Obs_Area_k1}  and \ref{def:Obs_Area_m1}. The challenge is to define global strategies that could lead to efficient solutions.

\begin{definition}[Global Observed Area]
Let $\boldsymbol{M}_{i}^{(j)}= (M_{i1}^{(j)},\ldots,M_{i\ell(i)}^{(j)})$, where $\ell(i)\in \{1,\ldots,k\}$ is the number of observed emitters in the $i$-th measurement that the $j$-th sensor gets. The global observed area is defined inductively, using the observed area of the individual sensor $P^{(s)}$ as follows:
\begin{align}
    P_1 = \bigcap_{s=1}^m \left( P^{(s)}_{1j} \right) = \bigcap_{s=1}^m \left( \bigcup_{j=1}^{\ell(1)} M^{(s)}_{1j} \right)
\end{align}
Let $P_{t-1}$ be the global observed area at time step $t-1$, and $\boldsymbol{M}^{(s)}_{t}$ be the current measurements of the $s$-th sensor. The global observed area at $t$, is $P_t$, where
\begin{align}
    P_t = \bigcap_{s=1}^m P_{t}^{(s)} = \bigcap_{s=1}^m \left(
    \bigcup_{j=1}^{\ell(t)} \left( P_{t-1}\bigcap M_{tj}^{(s)} \right)
    \right)
\end{align}
In this case the observed area can be interpreted as a union of polygons so with the notation $P_{ti}$ we shall refer to the observed area at time step $t$ and to the $i$-th polygon.
\end{definition}

\section{Study Cases}
\label{sec:Experiments}
A simulation environment modeling emitters and vehicles with sensors has been implemented to allow exploration of the approach and the resulting bounds.
In this section we will present experiments that reveal emergent behaviors of Algorithm~\ref{AlgCaseFrame}. 
Firstly it is helpful to show how the localisation works by starting with the case of ten emitters and four sensors who share their measurements in a shared memory storage. We run the simulation with the global objective~(\ref{eq:Go_max_step}) on the global observed area, and for $t_{max}=30$ steps.

\begin{figure}[ht]
    \centering
    \begin{tabular}{c c c}
        \scalebox{0.28}{
        \includegraphics{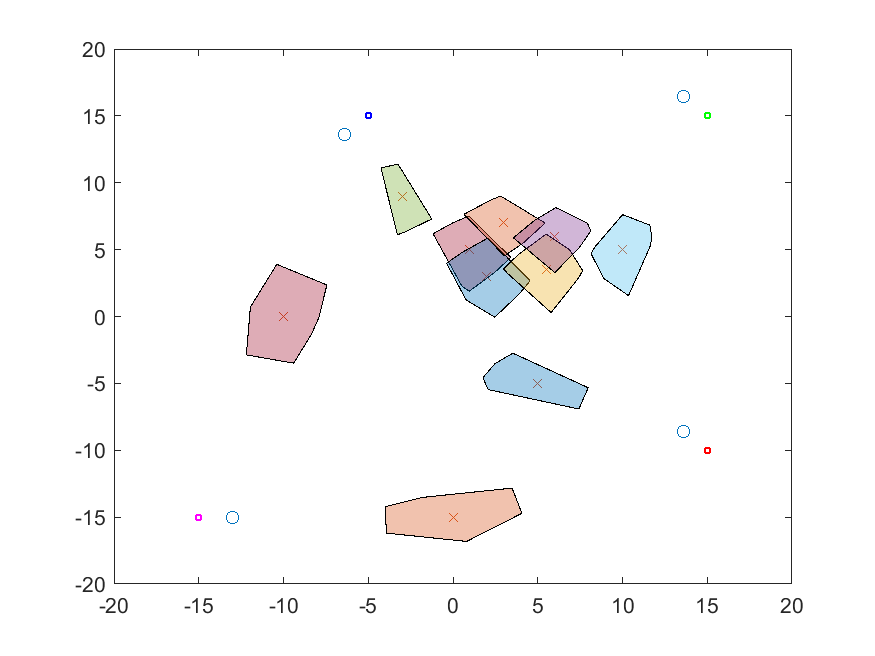}} & 
        \scalebox{0.28}{
        \includegraphics{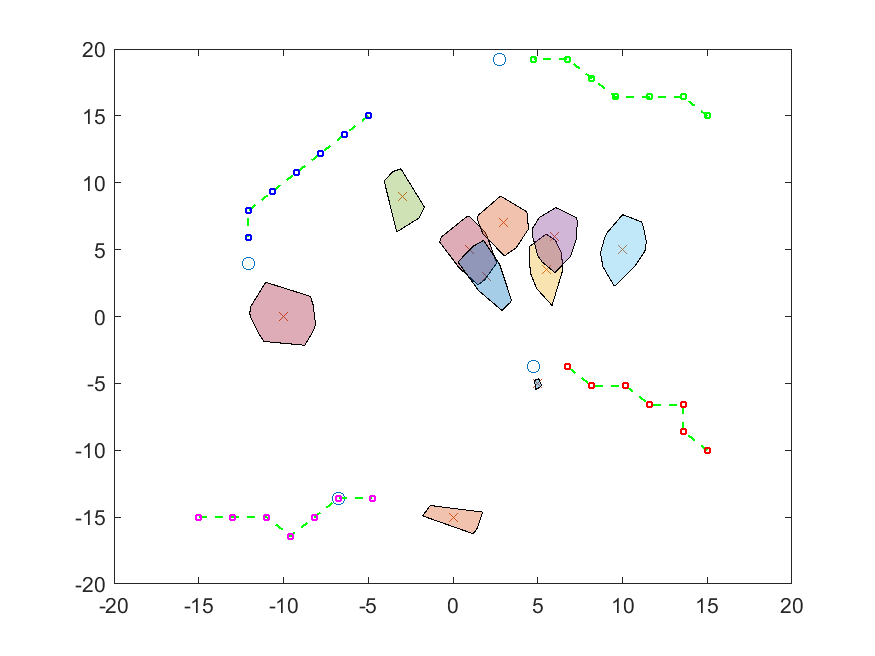}} &
        \scalebox{0.28}{
        \includegraphics{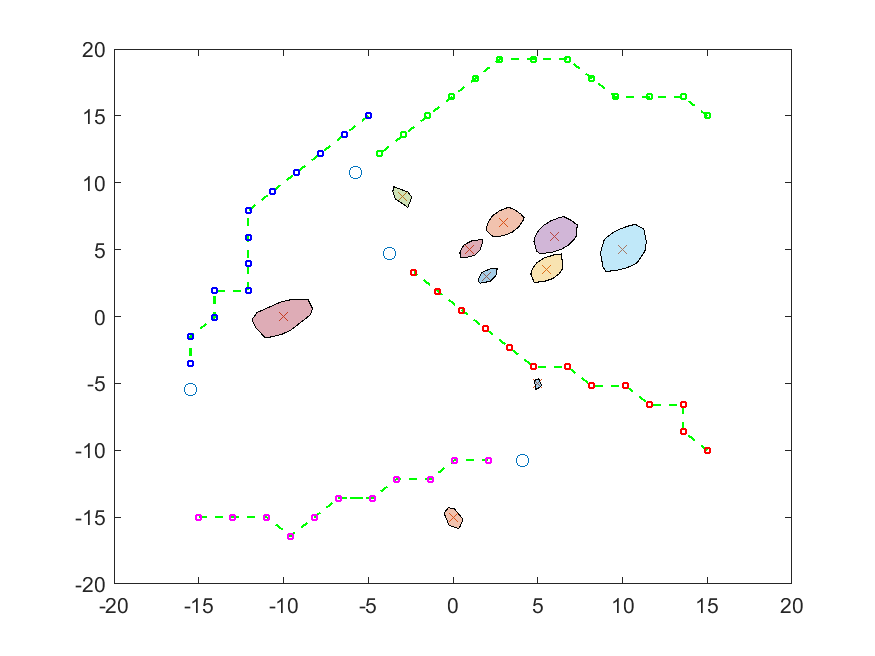}} 
        \\
        Step 2 & Step 7 & Step 13
        \\
        \scalebox{0.28}{
        \includegraphics{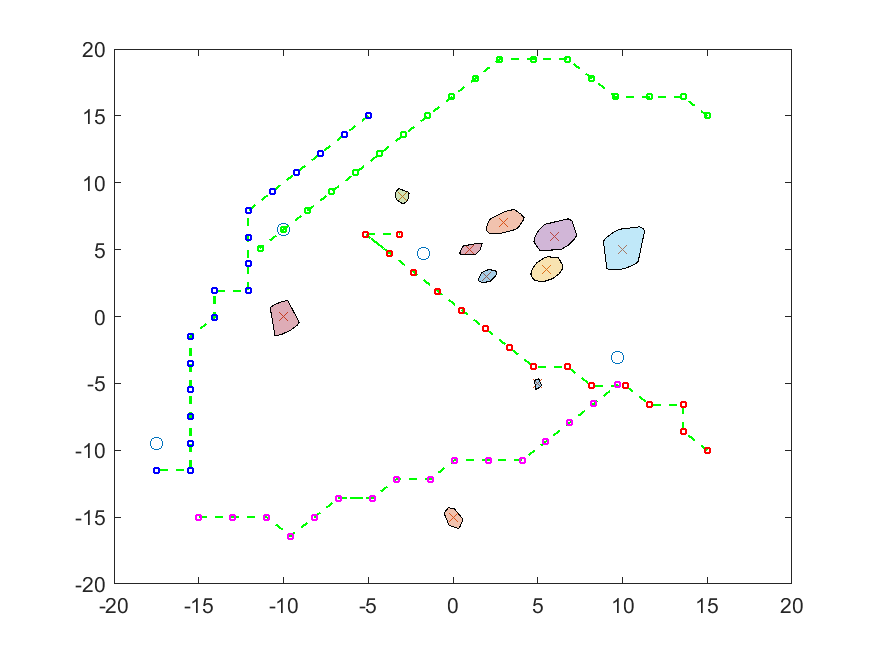}} 
        &
        \scalebox{0.28}{
        \includegraphics{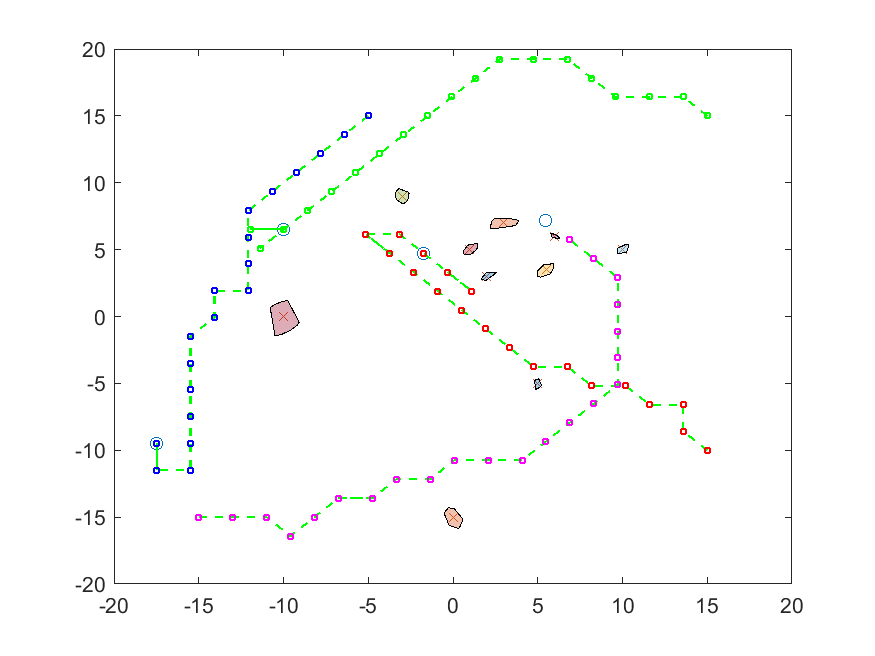}}
        &
        \scalebox{0.28}{
        \includegraphics{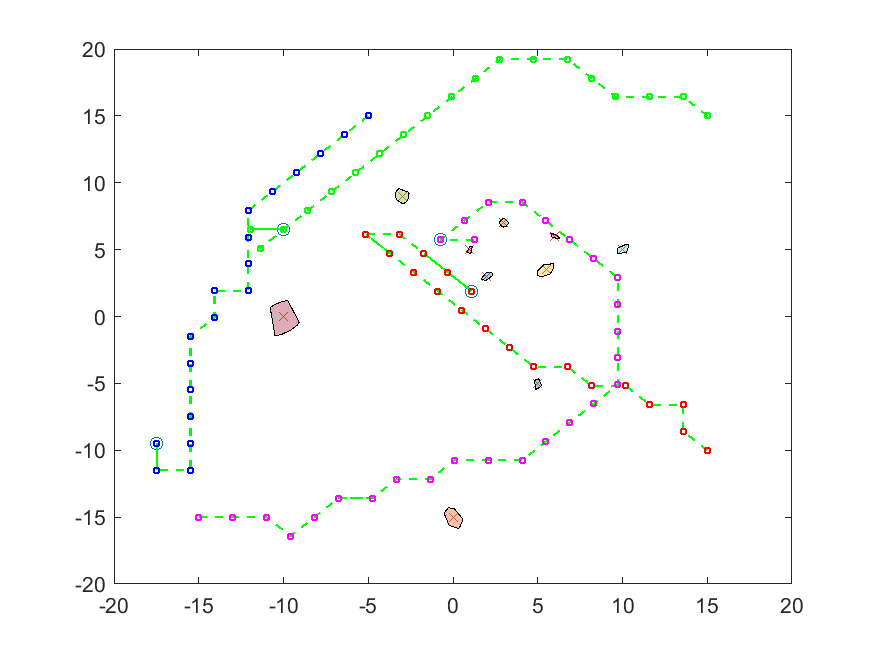}} 
        \\
        Step 18 & Step 24 & Step 30  
    \end{tabular}
    \caption{This is the run of the \textbf{greedy Algorithm with the objective function~(\ref{eq:Obj_area_1S_kE})}. With four sensors that share information but act independently.}
    \label{fig:Case_m_sen_k_emitters}
\end{figure}

In Figure~\ref{fig:Case_m_sen_k_emitters} the polygons shown are drawn from the shared memory that each sensor updates by intersecting the shared polygons with its partial information. But the sensors act independently which means they use their own partial information and the objective function~(\ref{eq:Obj_area_1S_kE}) to decide their next position.

One may observe in Figure~\ref{fig:Case_m_sen_k_emitters} between steps 24 and 30, trajectories seem to have stopped their expansion. This phenomenon happens because the greedy algorithm evaluates myopically,
by minimising the maximum uncertainty on the next step, and tends to oscillate between points inside a bounded area.

We have conducted experiments to get qualitative results by studying the emergent behaviour of the greedy approach between the objective functions (\ref{eq:Obj_area_1S_kE}) and~(\ref{eq:Obj_minmax_1S_kE}). 
Moreover, we examine how noise affects the greedy algorithm, and we study how taking into account more directions how affects the localisation. In the following experiments, we picked a representative realisation out of fifty in total to present the emergent behaviour.
%In the rest of this section we compare the greedy approach between the objective functions~(\ref{eq:Obj_area_1S_kE}) and~(\ref{eq:Obj_minmax_1S_kE}). Firstly, we examine how noise affects the greedy algorithm with different objective functions. Then we talk about the emergent behaviours of the two algorithms, one with objective  functions~(\ref{eq:Obj_area_1S_kE})  and~(\ref{eq:Obj_minmax_1S_kE}), respectively. Finally, we study how taking into account more directions how affects the localisation.

\subsection{Noisy Measurements}
Notice that if we include the noisy measurements as part of the input, then Algorithm~\ref{AlgCaseFrame} becomes deterministic,which means that for a fixed extended input it should always output one trajectory. To present how small differences in the noise affect the algorithm, we need to talk about how we represent the noise first.  We consider that if a sensor's field of view is $\phi$, then a measurement's maximum angular error is less than that. 
To implement this in the simulation, we perturb the line between a sensor and an emitter with Gaussian white noise. In reality, it does not matter if it is white noise as long as the perturbed line's angle does not exceed by $\phi/2$ the original line between the sensor and the emitter. 
\begin{figure}[ht]
    \centering
    \begin{tabular}{c c c c}
        \scalebox{0.22}{
        \includegraphics{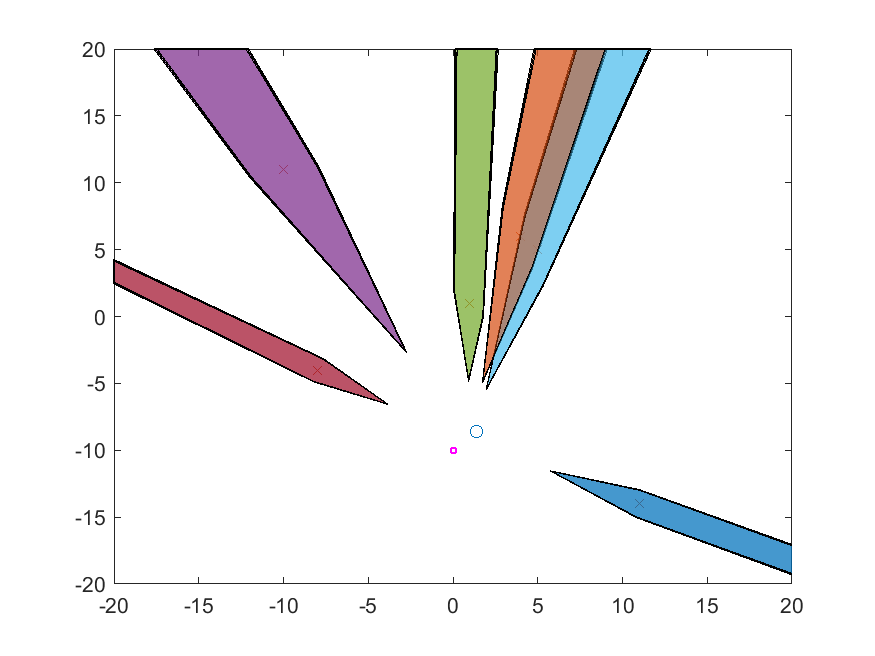}} & 
        \scalebox{0.22}{
        \includegraphics{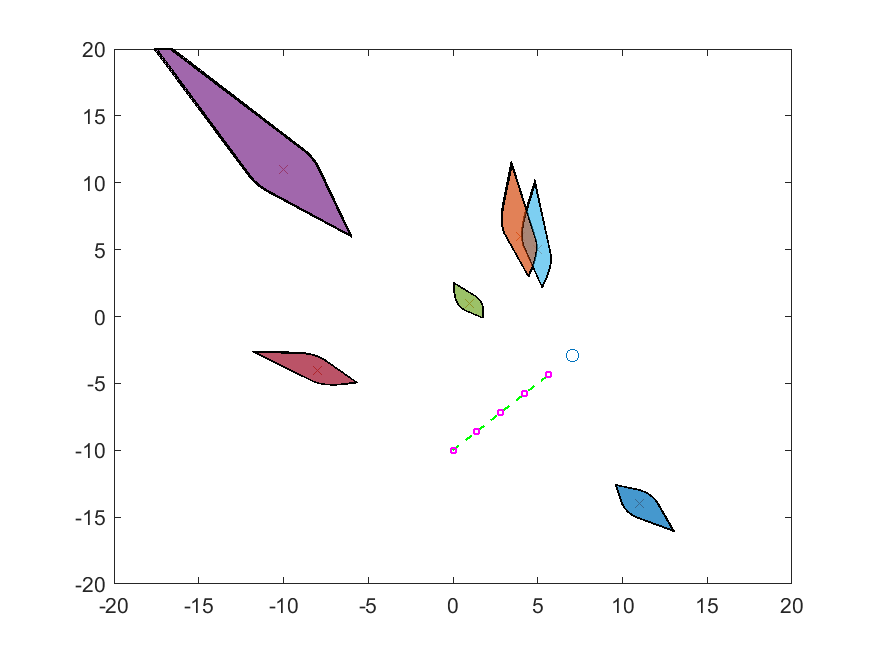}} &
        \scalebox{0.22}{
        \includegraphics{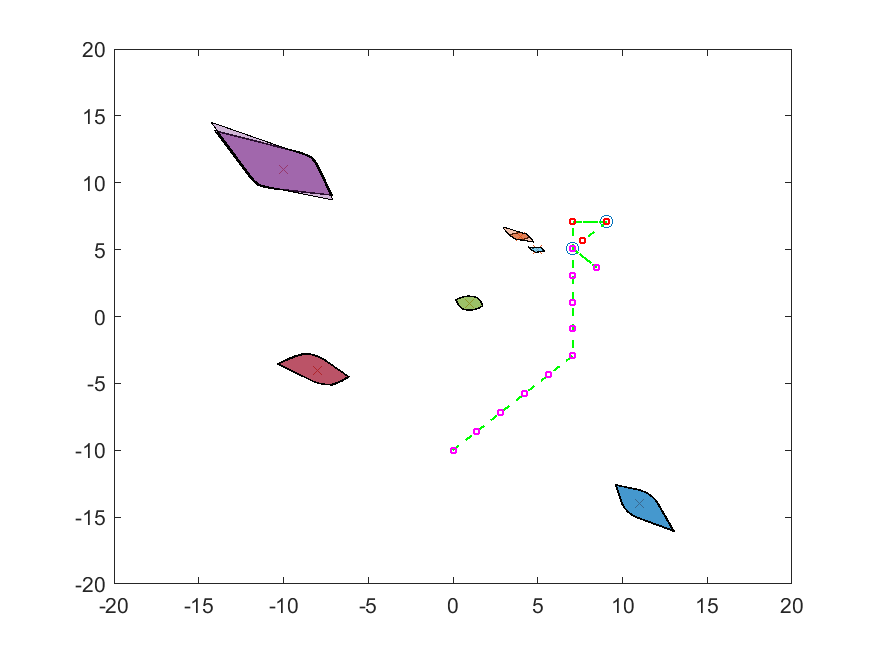}} &
        \scalebox{0.22}{
        \includegraphics{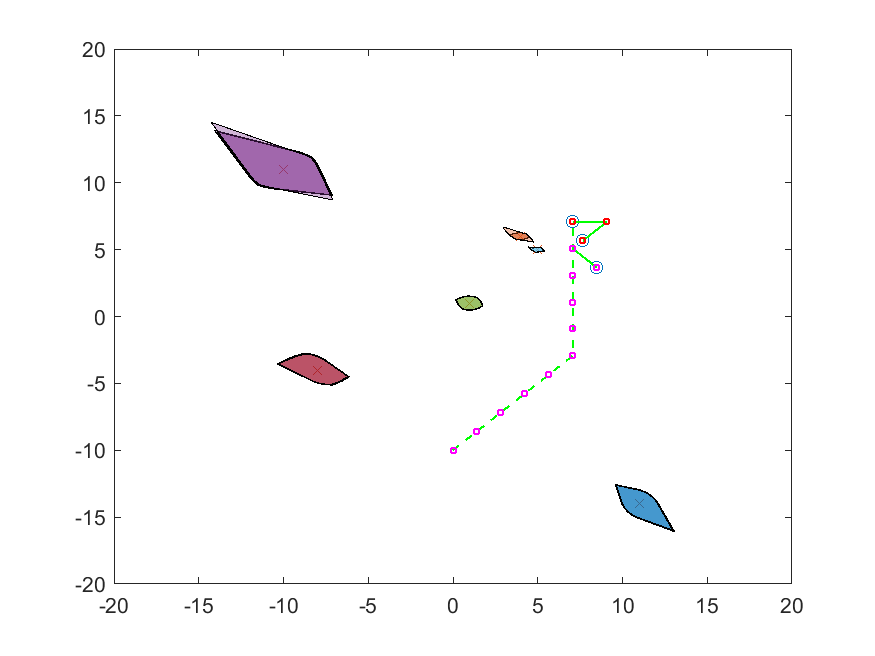}} 
        \\
        Step 2 & Step 5 & Step 13 & Step 20  
        %\\
        %\scalebox{0.26}{
        %\includegraphics{Images/sim1_step_4.7.png}} & 
        %\scalebox{0.26}{
        %\includegraphics{Images/sim1_step_4.11.png}} 
        %\\
        %Step 7 & Step 11
    \end{tabular}
    \caption{This is the run of the \textbf{greedy Algorithm with the objective function~(\ref{eq:Obj_area_1S_kE})}. In this one two different tracks have been created with small differences with one another.The polygons are drawn from the observed view of the individual sensor.}
    \label{fig:avg_noise}
\end{figure}
\begin{figure}[ht]
    \centering
    \begin{tabular}{c c c c}
        \scalebox{0.22}{
        \includegraphics{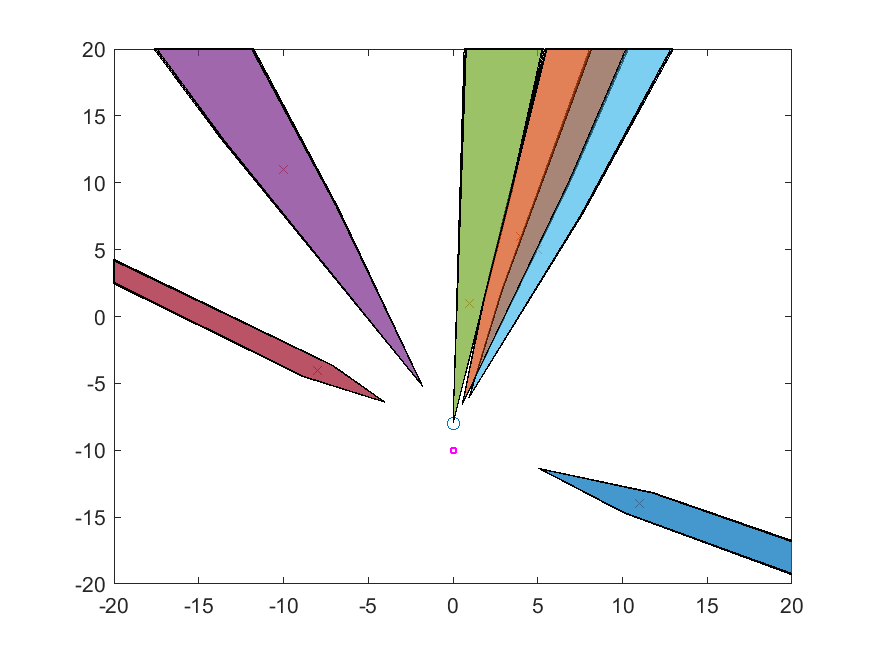}} & 
        \scalebox{0.22}{
        \includegraphics{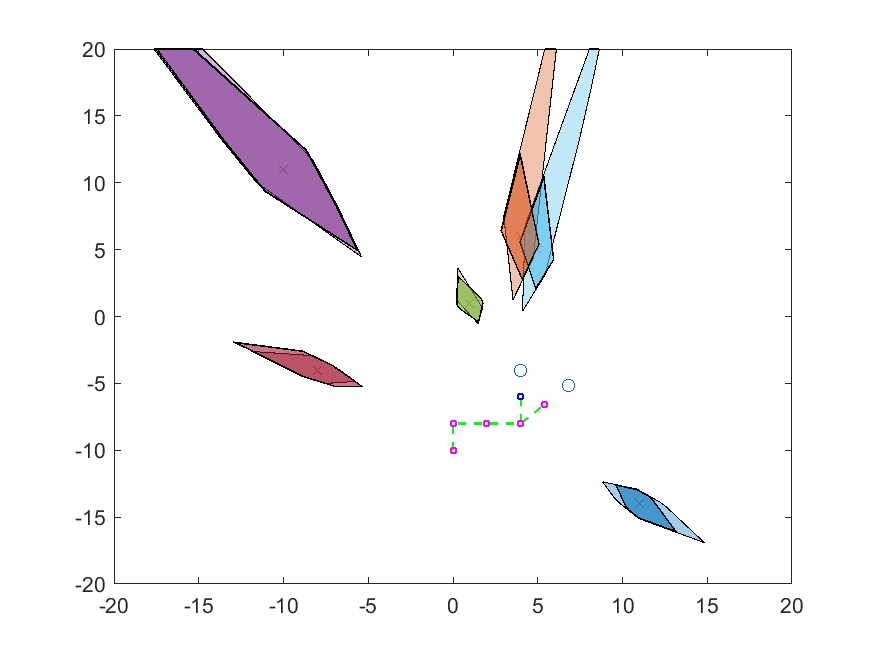}} &
        \scalebox{0.22}{
        \includegraphics{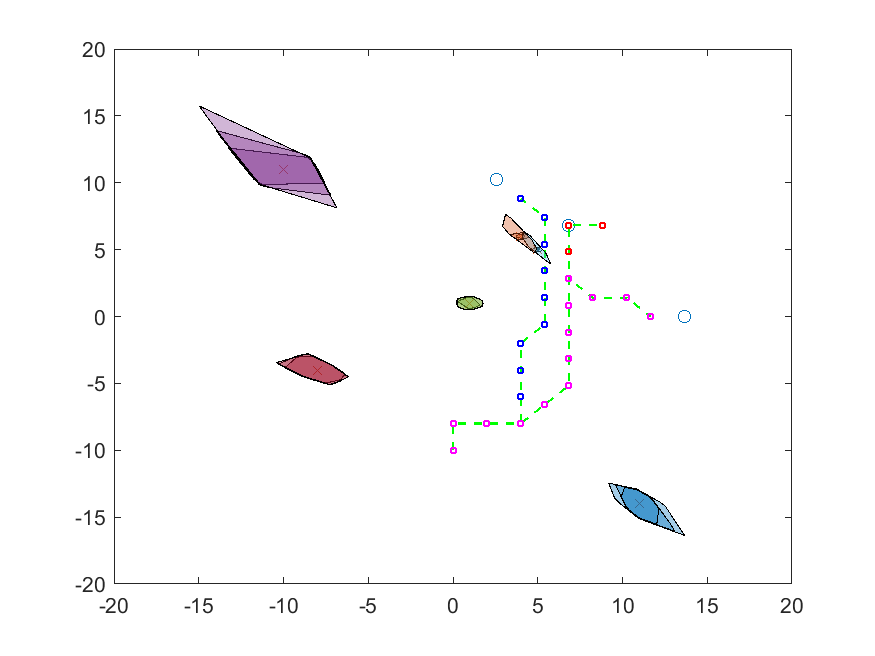}}
        &
        \scalebox{0.22}{
        \includegraphics{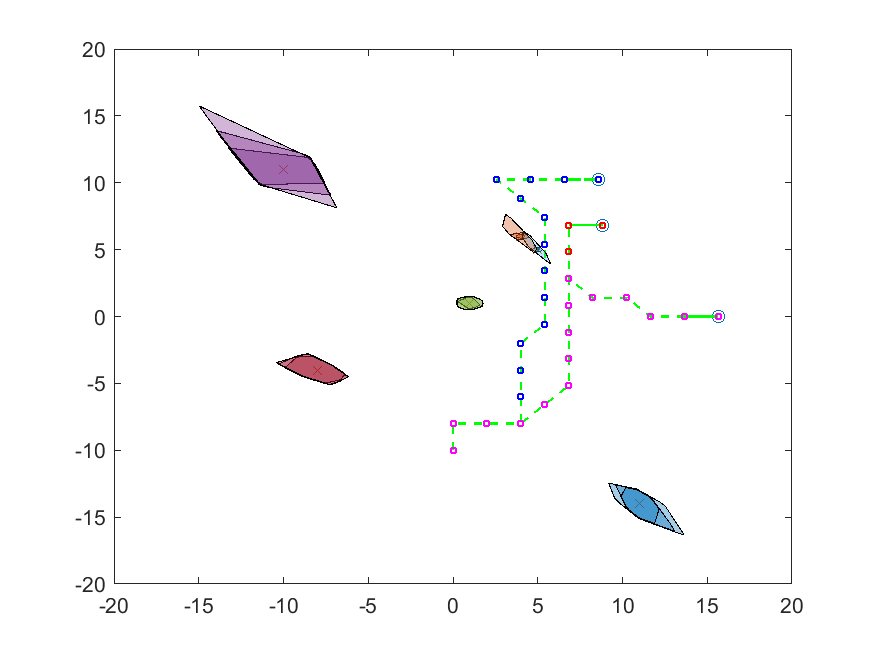}} 
        \\
        Step 2 & Step 5 & Step 11 & Step 20  
        %\\
        %\scalebox{0.26}{
        %\includegraphics{Images/MinMax_sim1_step_3.7.png}} & 
        %\scalebox{0.26}{
        %\includegraphics{Images/MinMax_sim1_step_3.11.png}} 
        %\\
        %Step 7 & Step 11
    \end{tabular}
    \caption{This is the run of the g\textbf{reedy Algorithm with the objective function~(\ref{eq:Obj_minmax_1S_kE})}. In this one three different tracks have been created because of the small differences on the level of noise. The polygons are drawn from the observed view of the individual sensor.}
    \label{fig:min_max_noise}
\end{figure}

In Figures~\ref{fig:avg_noise} and~\ref{fig:min_max_noise}  there are two typical simulations of the greedy algorithm with objective function~(\ref{eq:Obj_area_1S_kE}) and~(\ref{eq:Obj_minmax_1S_kE}) respectively. There are six emitters and three sensors that move and act independently, and they have a slightly different field of view and level of noise on their measurements. But they have the same set of directions and the same starting positions.

We observe that the algorithm with the objective function~(\ref{eq:Obj_area_1S_kE}) that minimises the average area is robust to small levels of noise, keeping the same trajectory. Whereas the algorithm with the objective function~(\ref{eq:Obj_minmax_1S_kE}) that minimises the worst-case polygon algorithm is observed to be more vulnerable to “small” noise differences as the worst-case polygon may change in each step, making the sensor wander around.

\subsection{Emergent Behavior}
Emergent behaviour provides a helpful perspective to study multi-agent systems in various settings~\cite{DBLP:journals/istr/ChenN22/Emerg2,DBLP:books/sp/Wang22/Emerg1}. 
When we have a system with simple rules, we may not be able to predict the outcome of an initial state (unless we simulate it step by step), due to the combinatorial explosion of possible outcomes, such a problem is the game of life which is Turing complete~\cite{DBLP:books/sp/02/Rendell02/Game_of_life}. The simulations of Algorithm~\ref{AlgCaseFrame} have randomised elements, and even if there is a large number of possible trajectories, a few of them are more likely to appear as the output than others in multiple executions of the same initial state. Analysing the emergent properties, represented by the different solutions, allows one to extract general properties from complex phenomena.
\begin{figure}[ht]
    \centering
    \begin{tabular}{c c c c}
        %\scalebox{0.26}{
        %\includegraphics{Images/Emerg_Avg_step_3.1.png}} & 
        %\scalebox{0.26}{
        %\includegraphics{Images/Emerg_Avg_step_3.9.png}} 
        %\\
        %Step 2 & Step 9
        %\\
        \scalebox{0.22}{
        \includegraphics{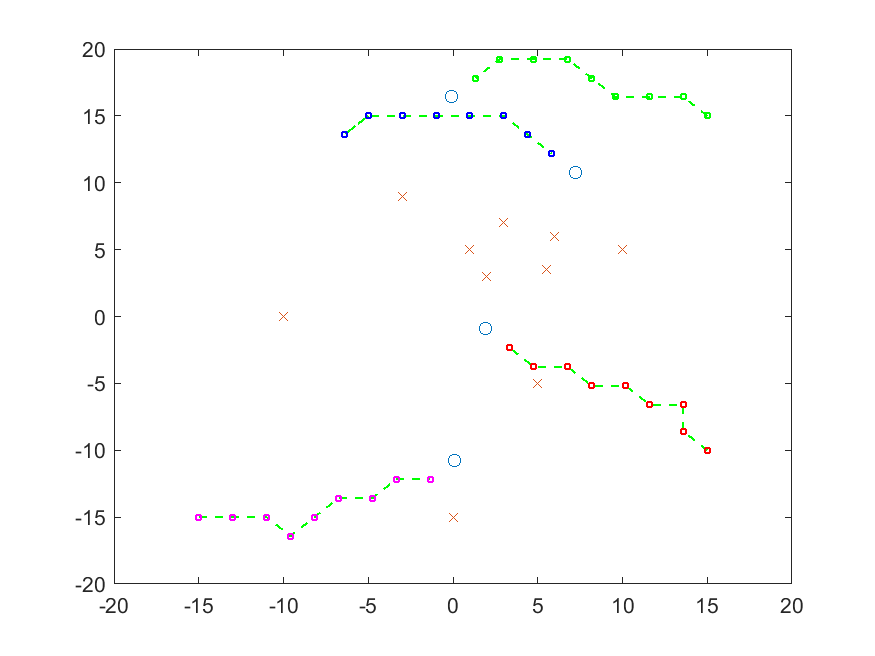}} & 
        \scalebox{0.22}{
        \includegraphics{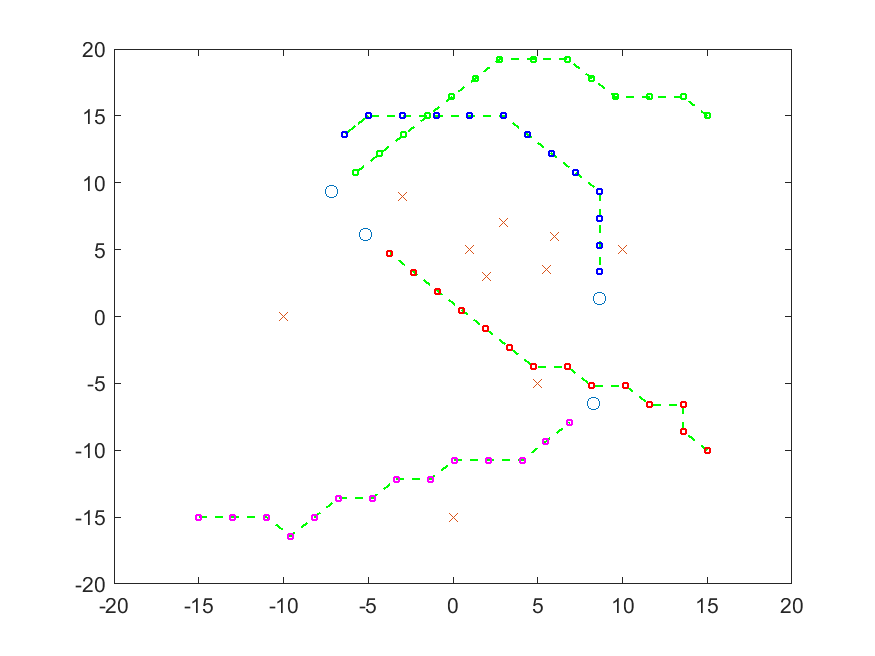}} &
        \scalebox{0.22}{
        \includegraphics{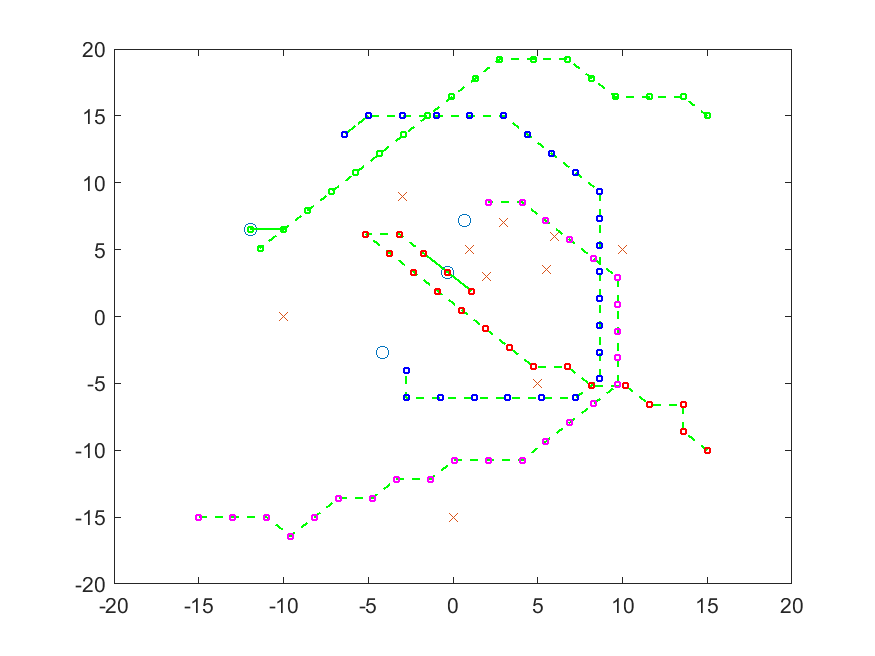}}
        &
        \scalebox{0.22}{
        \includegraphics{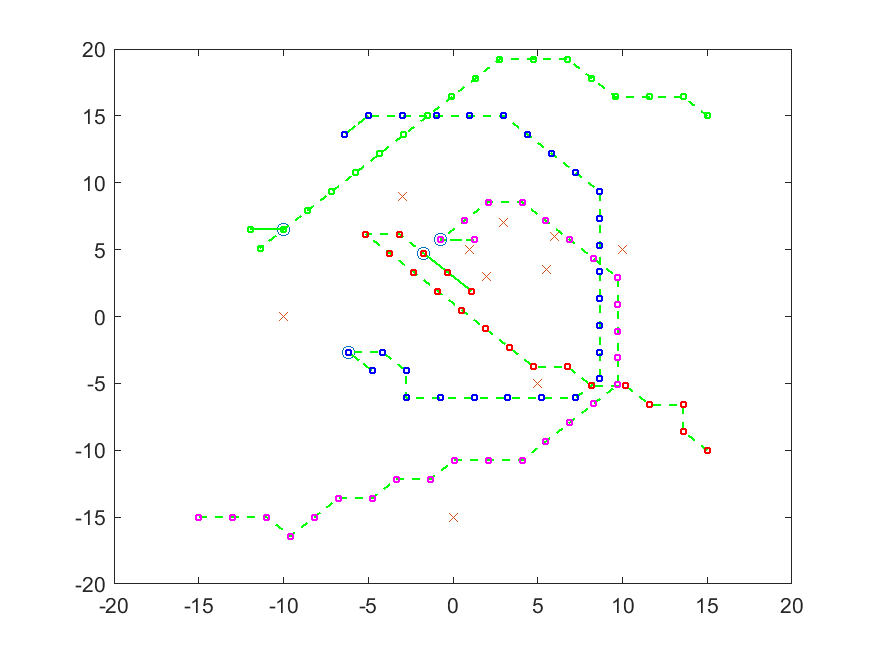}} 
        \\
        Step 9 & Step 14 & Step 25 & Step 30 
    \end{tabular}
    \caption{The run of the \textbf{greedy Algorithm with the objective function~(\ref{eq:Obj_area_1S_kE})}. There are four sensors, each one acting independently, and they are gravitated towards the cluster of the emitters in middle of the area.}
    \label{fig:avg_Emerg}
\end{figure}
\begin{figure}[ht]
    \centering
    \begin{tabular}{c c c c}
        %\scalebox{0.26}{
        %\includegraphics{Images/Emerg_minmax_step_1.0.png}} & 
        %\scalebox{0.26}{
        %\includegraphics{Images/Emerg_minmax_step_1.8.png}} 
        %\\
        %Step 2 & Step 8
        %\\
        \scalebox{0.22}{
        \includegraphics{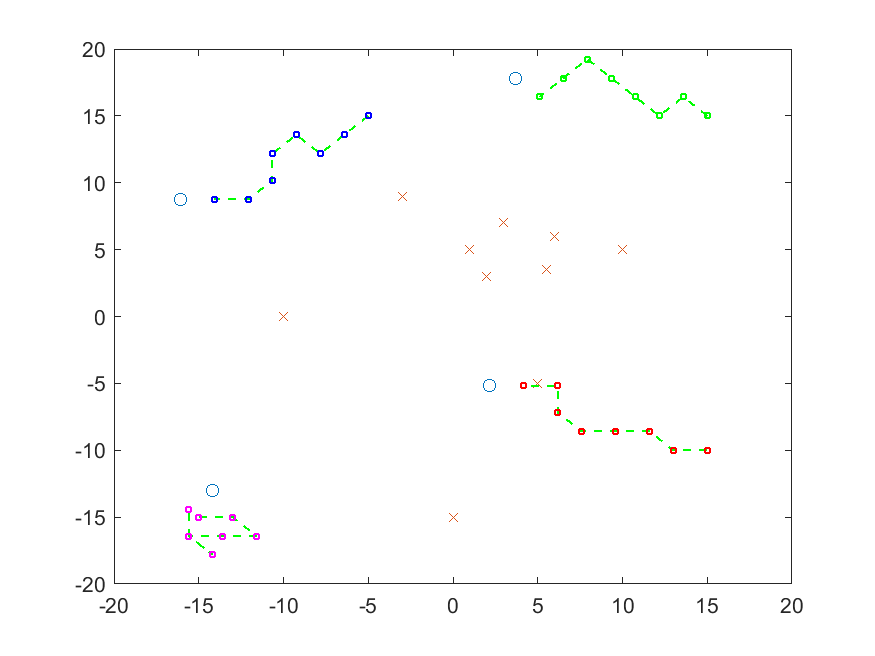}} & 
        \scalebox{0.22}{
        \includegraphics{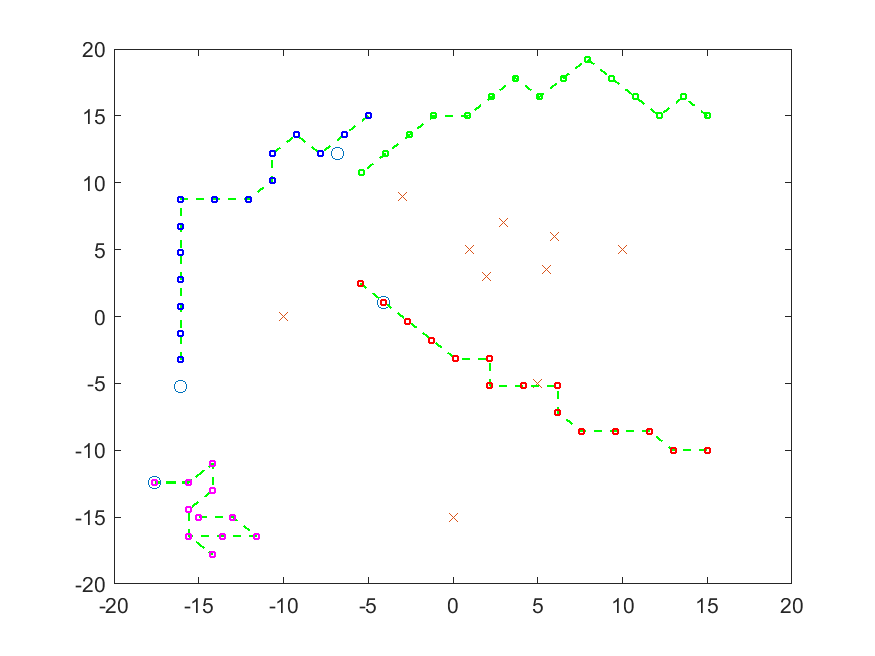}} 
        &
        \scalebox{0.22}{
        \includegraphics{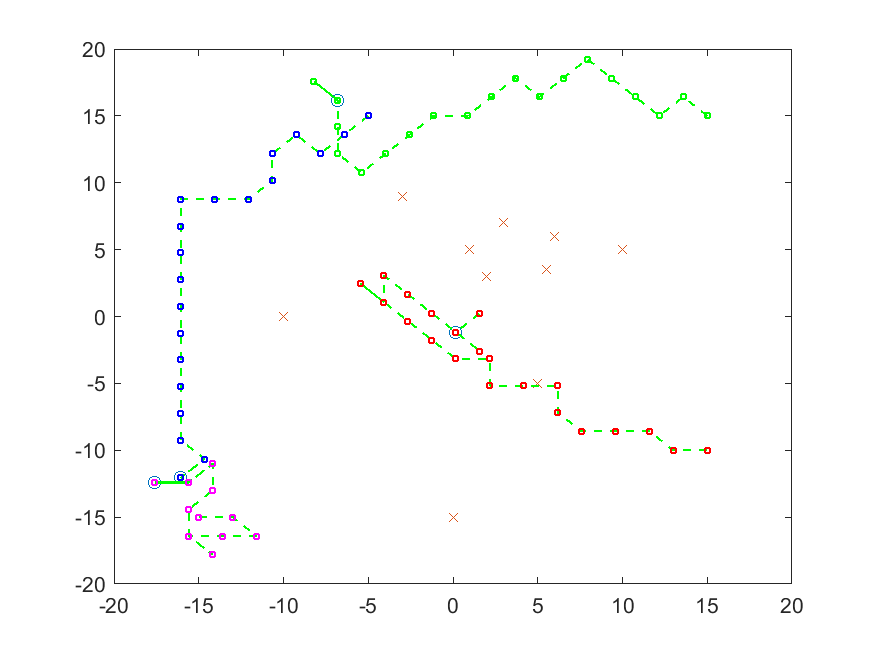}}
        &
        \scalebox{0.22}{
        \includegraphics{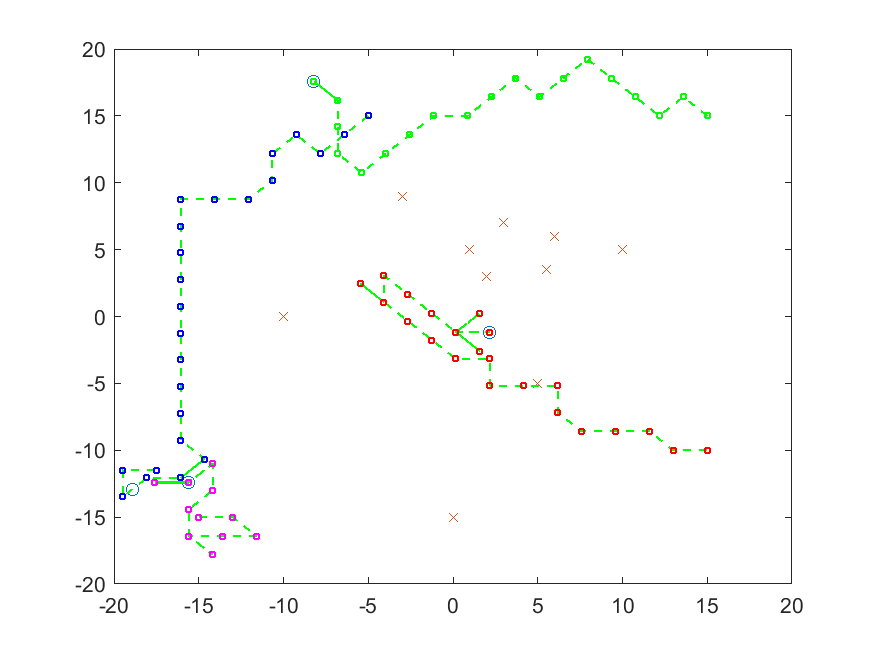}} 
        \\
        Step 8 & Step 15 & Step 23 & Step 30 
    \end{tabular}
    \caption{The run of the \textbf{greedy Algorithm with the objective function~(\ref{eq:Obj_minmax_1S_kE})}. There are four sensors, each one acting independently, and they appear to have a random movement.}
    \label{fig:minmax_Emerg}
\end{figure}

In Figures~\ref{fig:avg_Emerg} and~\ref{fig:minmax_Emerg} we present a simulation with four sensors that have the same field of view, the same set of possible directions, and objectives functions~(\ref{eq:Obj_area_1S_kE}) and~(\ref{eq:Obj_minmax_1S_kE}) respectively. The sensors act independently without sharing information. It becomes apparent that the sensors in figure~\ref{fig:avg_Emerg} gravitate towards the centre of the area, where a cluster of emitters lies, which appears to be the emergent behaviour of Algorithm~\ref{AlgCaseFrame} with objective function~(\ref{eq:Obj_area_1S_kE}).
But in Figure~\ref{fig:minmax_Emerg} the sensors' movement seems random, and it depends on the sensors' starting positions.

\subsection{Limiting the Sensor Motion}
\begin{figure}[ht]
    \centering
    \begin{tabular}{c c c c}
        \scalebox{0.22}{
        \includegraphics{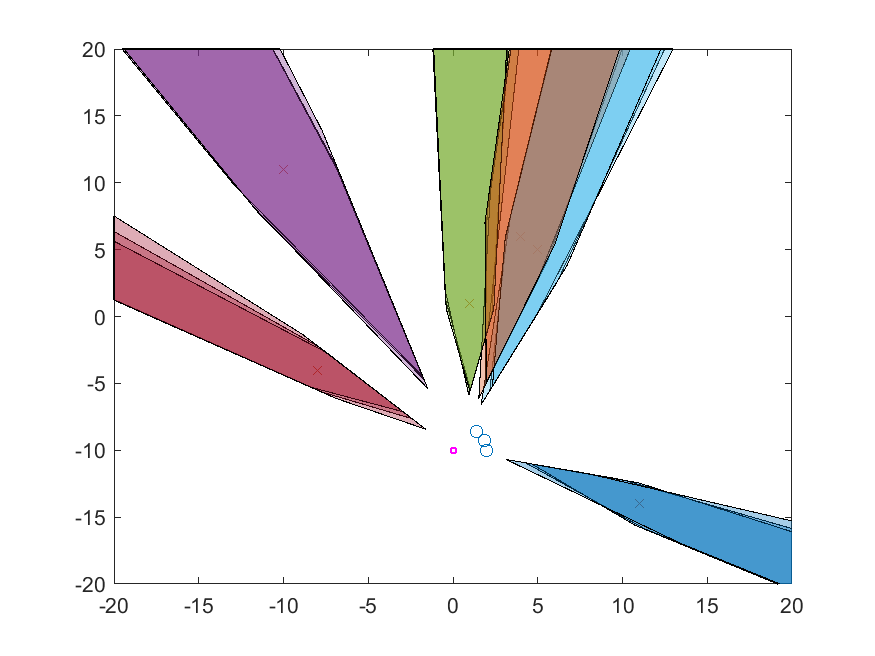}} & 
        \scalebox{0.22}{
        \includegraphics{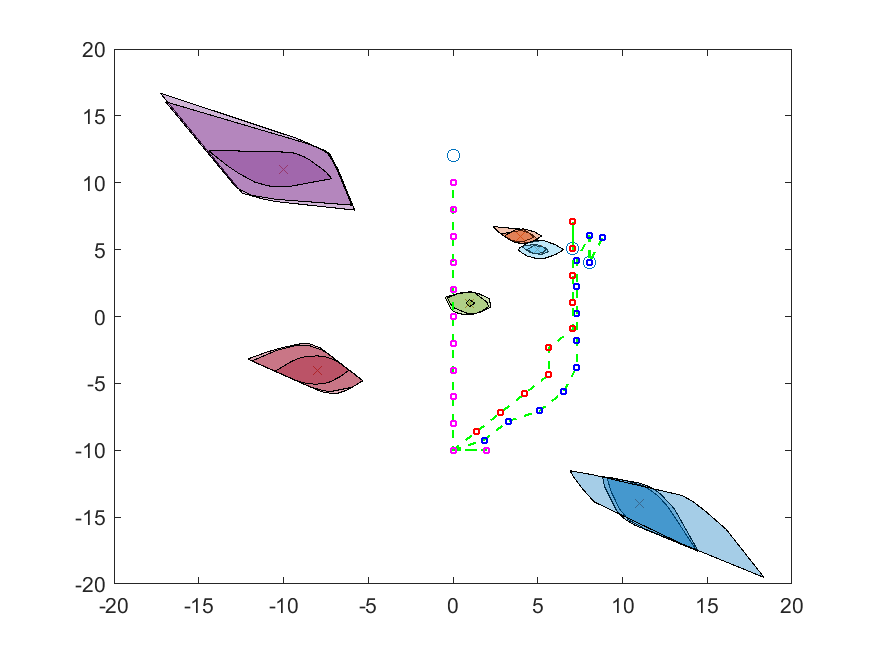}} &
        \scalebox{0.22}{
        \includegraphics{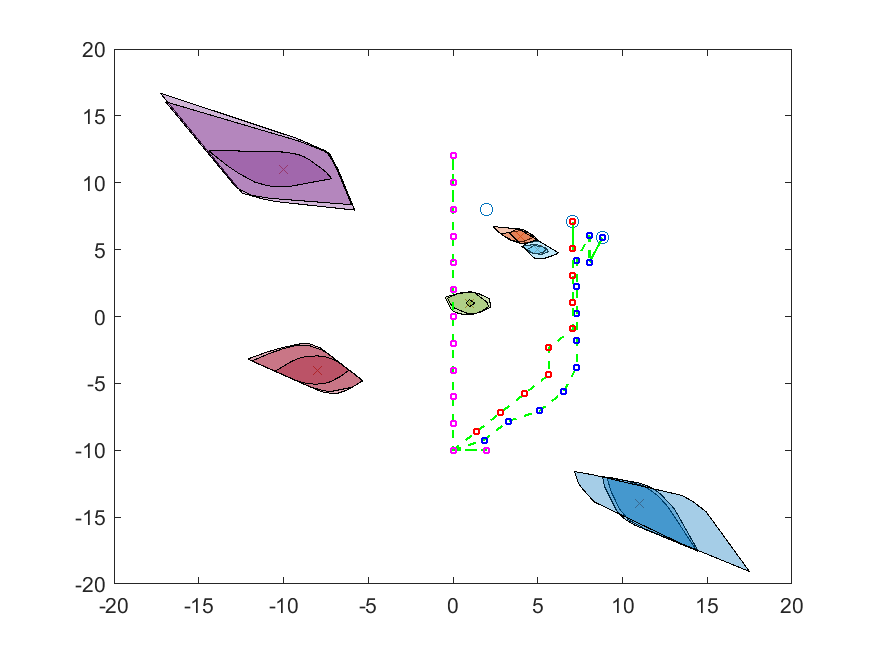}}
        &
        \scalebox{0.22}{
        \includegraphics{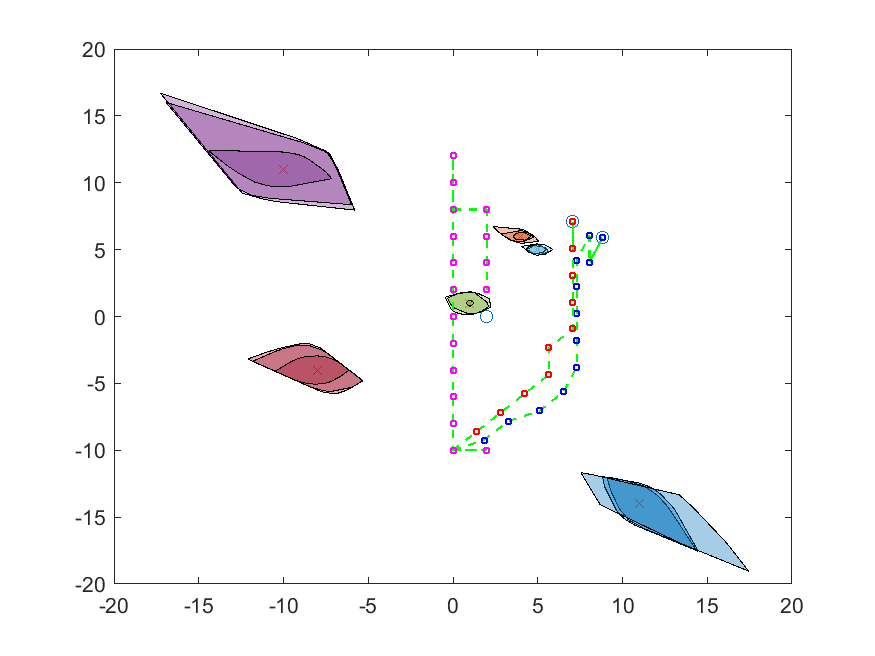}} 
        \\
        Step 2 & Step 13 & Step 16 & Step 20  
    \end{tabular}
    \caption{The simulation of the \textbf{greedy algorithm with objective function~(\ref{eq:Obj_area_1S_kE})}. The three sensors, blue, red, and purple have 16, 8, and 4 directions available as they move.  }
    \label{fig:Directions_avg}
\end{figure}

\begin{figure}[ht]
    \centering
    \begin{tabular}{c c c c}
        \scalebox{0.22}{
        \includegraphics{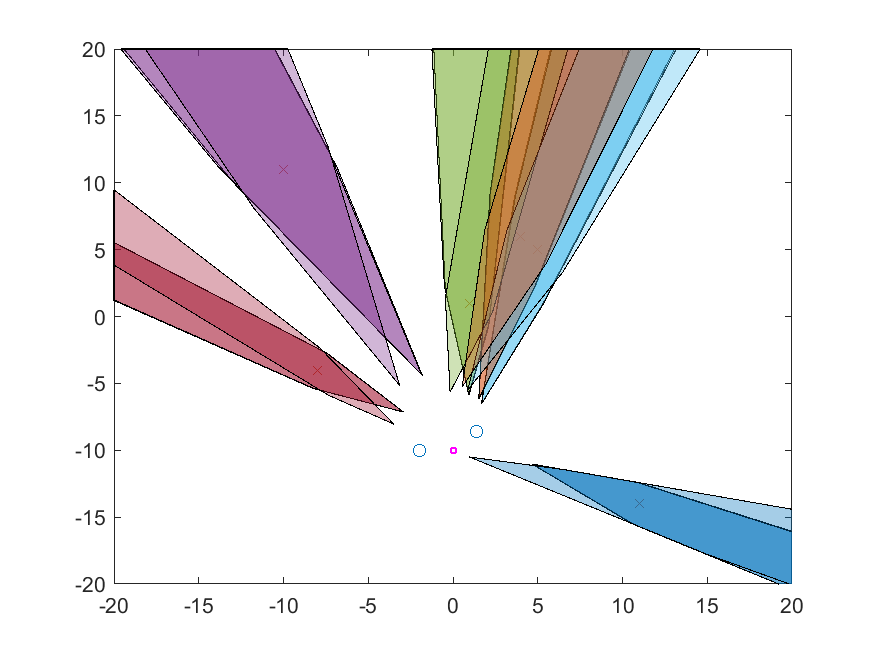}} & 
        \scalebox{0.22}{
        \includegraphics{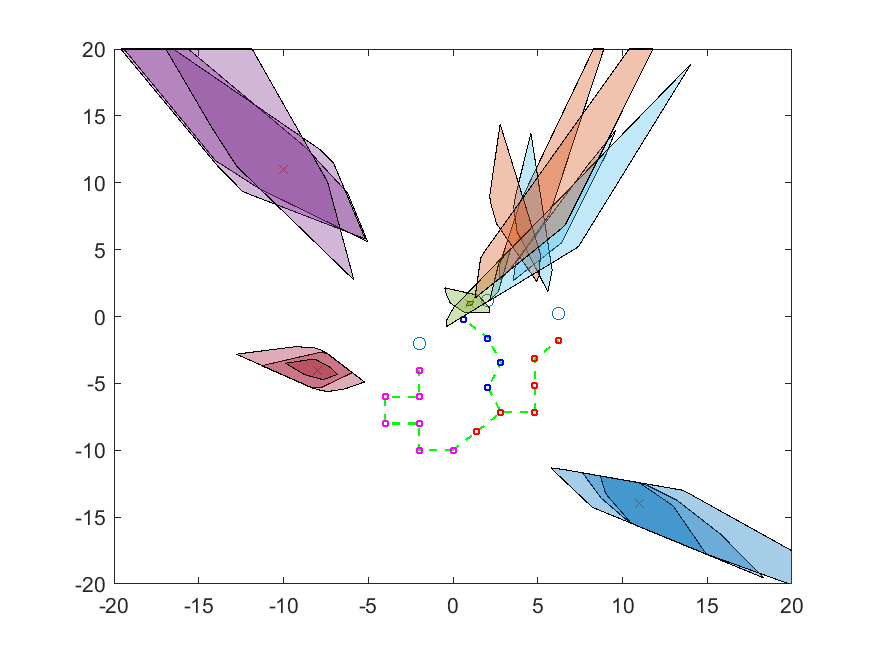}} 
        &
        \scalebox{0.22}{
        \includegraphics{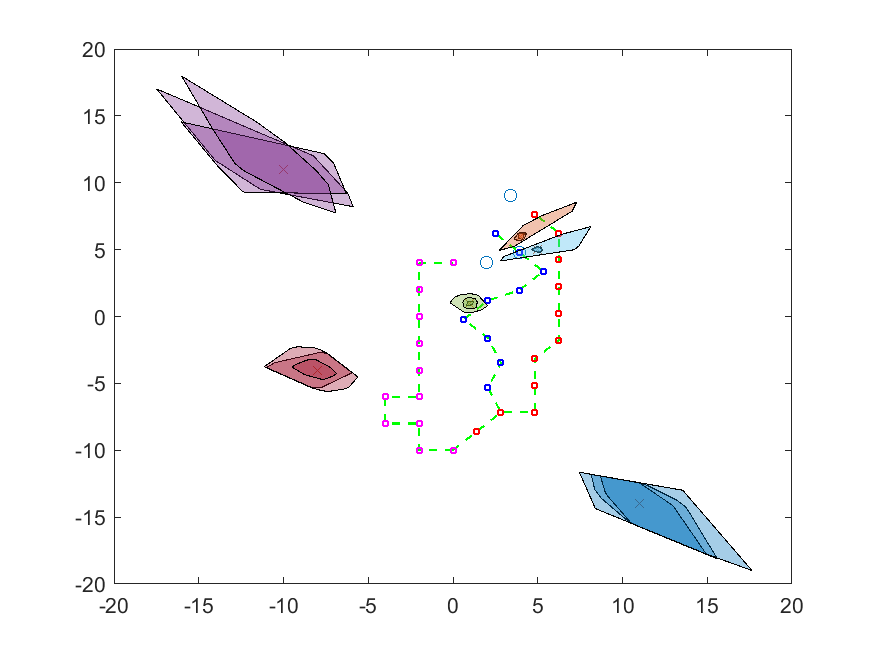}}
        &
        \scalebox{0.22}{
        \includegraphics{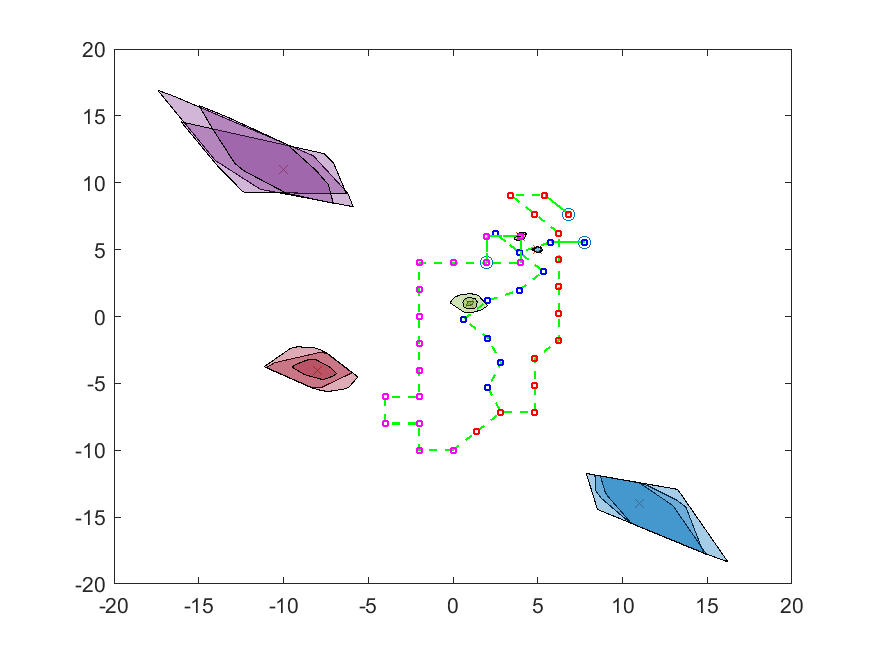}} 
        \\
        Step 2 & Step 7 & Step 12 & Step 20 
    \end{tabular}
    \caption{The simulation of \textbf{the greedy algorithm with objective function~(\ref{eq:Obj_minmax_1S_kE})}. The three sensors, blue, red, and purple have 16, 8, and 4 directions available as they move. }
    \label{fig:Directions_minmax}
\end{figure}

Another parameter worth examining is the number of directions that a sensor is allowed to move. In this simulation, we have three sensors with the same field of view, starting position, and objective function. Moreover, they act independently without sharing information. The only difference is the set of available directions. We use $D_1,D_2,D_3$ with $D_3\subseteq D_2\subseteq D_1$, $|D_1|=16$, $|D_2|=8$, and $|D_3|=4$. In Figures~\ref{fig:Directions_avg} and~\ref{fig:Directions_minmax} we have the simulations of the greedy algorithms with objective functions~(\ref{eq:Obj_area_1S_kE}), and~(\ref{eq:Obj_minmax_1S_kE}) respectively. 

\begin{comment}
\begin{figure}[ht]
    \centering
    \begin{tabular}{c c}
        \scalebox{0.16}{
        \includegraphics{Images/plot_avg_area_No_Directions_Avg_Algo_4.jpg}} 
        &
        \scalebox{0.16}{
        \includegraphics{Images/plot_avg_area_No_Directions_MinMax_Algo_3.jpg}}
        \\
         (a) & (b) 
    \end{tabular}
    \caption{The average area over the number of steps. In (a) and (b) we have the greedy algorithm with objective function~(\ref{eq:Obj_area_1S_kE}) and~(\ref{eq:Obj_minmax_1S_kE}) respectively.}
    \label{fig:plot_area_direction}
\end{figure}
\end{comment}

\begin{figure}[ht]
    \centering
    \begin{tabular}{c c}
        \scalebox{0.38}{
        \includegraphics{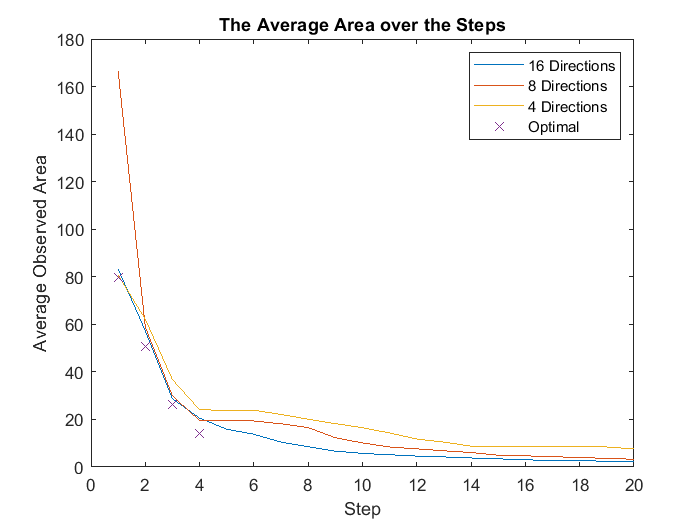}} &  
        \scalebox{0.38}{
        \includegraphics{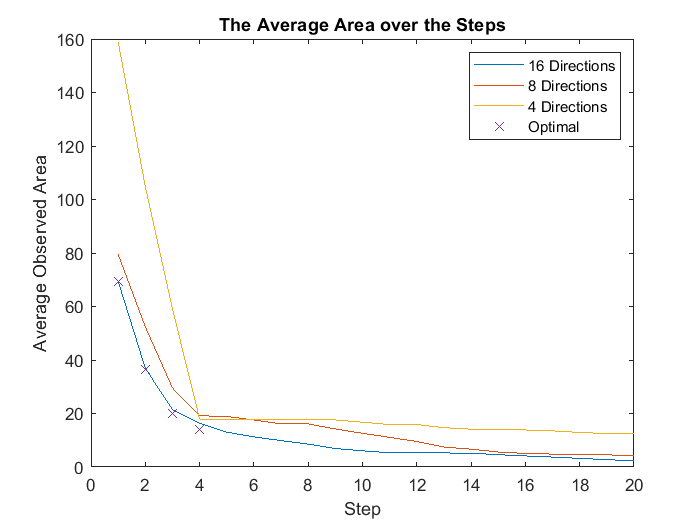}}
        \\
        (a) & (b)
    \end{tabular}
    \caption{(a) The average area over the number of steps when using the \textbf{greedy algorithm with objective function~(\ref{eq:Obj_area_1S_kE})}. (b) The average area over the number of steps when using the \textbf{greedy algorithm with objective function~(\ref{eq:Obj_minmax_1S_kE})} }
    \label{fig:plot_avg_nodirections}
\end{figure}

From the Figure~\ref{fig:plot_avg_nodirections} we observe that the more the directions the faster the localisation of the greedy algorithm. Moreover, with crosses we have marked the optimal observed area by running an exhaustive algorithm on 16 available directions.
But we should also mention that there are instances where the sensor with less direction choices achieves a better result which the minimisation of the observed average area.

\section{Discussion}
\label{sec:Discuss}
The problem of localising the emitters' positions with passive sensors is computationally intractable
%infeasible
if we consider all of the sensors' possible trajectories. Our approach provides a solution which is close to optimal without computationally expensive procedures. The computed polygons are reliable bounds on the emitters' location. Moreover they are obtained only using the geometry of the problem, without assuming a probability or a belief distribution. This new approach offers a fresh perspective to consider emergent behaviours and high-level planning of global strategies to achieve given goals.

The greedy algorithm gives an estimate for every possible next position according to the selected objective function. We have observed that the algorithm works well, until it reaches a point where it moves inside a bounded area where it oscillates between points that have the minimum estimation but they are offering little new information overall.

An emergent behaviour that arises from the greedy algorithm with the objective function~(\ref{eq:Obj_area_1S_kE}) is that the sensor gravitates towards the largest cluster of emitters until it converges to a point where it moves to a bounded area, oscillating and gaining little new information. Nevertheless, the greedy algorithm with the objective function~(\ref{eq:Obj_minmax_1S_kE}) appears to be more erratic, as it is affected by the maximum uncertainty of only one polygon in each time step. In other words, objective function~(\ref{eq:Obj_minmax_1S_kE}) is prone change the focus of the sensor in every time step, creating back and forth movements, which makes the localisation process slower.

Moreover, a greedy algorithm with objective function~(\ref{eq:Obj_area_1S_kE}) that considers the average uncertainty among the polygons is more robust to small changes in the noise than its counter part with objective function~(\ref{eq:Obj_minmax_1S_kE}) that considers the uncertainty of just one polygon. Finally, one interesting fact is that, using more directions available to the sensor's movement makes localisation faster. However, there are cases when the sensors' movement is restricted, and the sensor still reaches a marginally better localisation.

%For future work, the centralised approach presented here is closely related to the problem of gossiping. 
%An interesting 
For future work, an interesting
idea would be to study this problem in temporal graphs as the sensors move and may get out of communication range. In addition, it would be interesting to study the decision-making mechanisms presented in this paper in a distributed rather than a centralised environment.

\newpage

\bibliographystyle{unsrt}
\bibliography{bibliography.bib}

\end{document}